\documentclass[10pt]{asme2ej}
\usepackage{appendix}
\usepackage{longtable}
\usepackage{epsfig} 
\usepackage{amsmath}
\usepackage{multicol}
\usepackage{fancyhdr}
\usepackage{lipsum}

\title{Thermal Model of an Omnimagnet for Performance Assessment and \\ Temperature Control\thanks{A preliminary version of this work was presented in IMECE 2017: paper number IMECE2017-72120}}
\usepackage{setspace}
\author{Fateme Esmailie
    \affiliation{Graduate Research Assistant\\
	Dept. of Mechanical Engineering\\
	University of Utah\\
	Salt Lake City, UT, 84112\\
	Email: ameel@mech.utah.edu 
    }	
}
\author{Matthew S. Cavilla
    \affiliation{ Graduate Research Assistant\\
	Dept. of Mechanical Engineering\\
	University of Utah\\
	Salt Lake City, UT, USA\\
	Email: Matt.cavilla@utah.edu  
    }
}
\author{Jake J. Abbott\\ Professor \\ 
Dept. of Mechanical Engineering\\
        University of Utah\\ Salt Lake City, UT, USA\\
        Email: jake.abbott@utah.edu\\
        
       {\tensfb Tim A. Ameel}\thanks{Address all correspondence for other issues to this author.} 
        \affiliation{Professor \\ Dept. of Mechanical Engineering\\
	University of Utah\\
	Salt Lake City, UT, USA\\
	Email: ameel@mech.utah.edu 
    }
    
}

\begin{document}

\maketitle    
\newpage
\begin{abstract}
{\it 
An Omnimagnet is an electromagnetic device that enables remote magnetic manipulation of devices such as medical implants and microrobots. It is comprised of three orthogonal nested solenoids with a ferromagnetic core at the center. Electrical current within the solenoids leads to Joule heating, resulting in undesired temperature increase within the Omnimagnet. If the temperature exceeds the melting point of the wire insulation, device failure will occur. Thus, a study of heat transfer within an Omnimagnet is a necessity, particularly to maximize the performance of the device. For the first time, a transient heat transfer model, that incorporates all three heat transfer modes, is proposed and validated with experimental data for an Omnimagnet with maximum root mean square error equal to 8${\%}$ (4$^{\circ}$C). This transient model is not computationally expensive. It is relatively easy to apply to Omnimagnets with different structures. The accuracy of this model depends on the accuracy of the input data. The code is applied to calculate the maximum safe operational time at a fixed input current or the maximum safe input current for a fixed time interval. The maximum safe operational time and maximum safe input current depend on size and structure of the Omnimagnet and the lowest melting point of all the Omnimagnet materials. A parametric study shows that increasing convective heat transfer during cooling, and during heating with low input currents, is an effective method to increase the maximum operational time of the Omnimagnet.The thermal model is also presented in a state-space equation format that can be used in a real-time Kalman filter current controller to avoid device failure due to excessive heating. }
\end{abstract}

\b{\it{Keywords}}: Transient; Lumped capacitance; State-space; Thermal management; Omnimagnet; Optimization


\pagenumbering{arabic}
\section{Introduction}

An Omnimagnet \cite{Petruska2014-2} is a relatively new electromagnetic device that enables remote magnetic manipulation \cite{Petruska2014} of devices such as medical implants and microrobots. We are particularly interested in its use for robotically assisted insertion of cochlear-implant electrode arrays \cite{Clark,Leon1, Leon2}, in which an Omnimagnet is adjacent to the patient's head during surgery. An Omnimagnet is comprised of three orthogonal nested solenoids with a spherical ferromagnetic core at the center, and is optimized to generate a dipole-like magnetic field in any direction. By controlling the input current within the individual solenoids, the magnitude and direction of the resulting dipole moment can be controlled. The electrical current flowing through the solenoids produces Joule heating, which results in undesirable temperature increase within the Omnimagnet. If the temperature of the wire insulation exceeds its melting point in any of the solenoids, a short circuit will occur, resulting in irreversible device failure. Thus, a heat transfer study is necessary to define operational limits for the Omnimagnet. Additionally, a heat transfer model can be employed to improve future designs of Omnimagnets. 

Petruska and Abbott \cite{Petruska2014-2} briefly considered heat transfer in an Omnimagnet as a part of the original design process. They assumed a steady-state condition and calculated the maximum current density for a desired magnetic field strength. Transient heat transfer within the Omnimagnet has not been previously studied. Other relevant research is related to heat-transfer studies within power transformers. The heat source in both a power transformer and an Omnimagnet is resistive heating and the materials (copper, electrically resistive paper, and polyamide insulation) are similar. Thus, the basic heat transfer within a power transformer and an Omnimagnet are alike; hence, the published work on power-transformer heat transfer is relevant. Temperature is a key factor in controlling power transformer aging. Based on the international standards for oil-immersed transformers (IEC 60076-7), the aging rate of power transformers is normal at temperatures lower than 98$^{\circ}$C. For every 6$^{\circ}$C temperature increase, the lifetime of a transformer is reduced by 50\% \cite{campelo1}. Another key performance parameter in transformers is hot-spot location. Locating hot spots and calculating their temperature are two of the primary reasons researchers have investigated power-transformer heat transfer \cite{GASTELURRUTI,HOSSEINI,MUFUTA,TSILI,ALLAHBAKHSHI}. 

Computational fluid dynamics (CFD) and thermal network models (TNM)/thermal hydraulic network models (THNM) are two of  the  main techniques that have been applied to model heat transfer within power transformers \cite{campelo1, TSILI,ALLAHBAKHSHI,SMOLKA,BEIZA,campelo2,CODDE,RODRIGUEZ,TORRIANO,Lecuna}. In the review paper by Campelo et al., it was reported that increasing oil flow rate does not significantly change the convection coefficient in high voltage power transformers \cite{campelo1}. They concluded that CFD and TNM methods produced the most accurate thermal models for power transformers. CFD methods were reported to be more accurate, but computationally expensive, and as a result researchers are using TNM frequently in their studies \cite{campelo2}.

Research studies focused on modeling heat transfer within power transformers are still being reported \cite{campelo1, ALLAHBAKHSHI,CODDE,RODRIGUEZ,Liu}. For instance, Rodriguez et al.\ \cite{RODRIGUEZ} focused on the cooling capacity of radiators that have been used to decrease the temperature of the windings.  They performed experiments to validate their simulation results and showed that oil is ten times more efficient than air in cooling the transformer. They also reported that the air flow rate has only a minor effect on convective cooling of the transformer.

In most heat-transfer studies related to power transformers, a steady-state condition has been applied. Power transformers are relatively large devices; thus, finding the spatial temperature distribution of the power transformers has been the main aim of most studies. On the contrary, an Omnimagnet is typically a relatively small device in comparison to a power transformer; therefore, the transient temperature change of the Omnimagnet is more important than the spatial temperature distribution. Because of the small dimensions of the Omnimagnet, assuming a uniform temperature distribution within individual components is reasonable, i.e., the Biot number ($Bi$) is small. Probably the most important determination in an Omnimagnet heat-transfer study is the maximum time the device can be operated, for a given power distribution in the three coils, before its failure. Alternatively, one is interested in the maximum power that can be applied for a certain period of time. Therefore, an assumption of a steady-state condition is not beneficial for an analysis of Omnimagnet thermal behavior.

Methods to mitigate temperature rise in electrical devices such as power transformers have also been pursued recently.  For instance, modified nanomaterials with better thermal properties (higher thermal conductivity and higher cooling convection coefficient) \cite{CONTRERA,Fontes,Guan} and magnetic fluid coolants \cite{Modestov,PATEL} are two interesting adaptations that have been studied. Mineral oil has been used as a cooling fluid in transformers. Nanofluids have also been used as alternative coolants. The thermal conductivity of nanofluids can be as much as 75\% greater than that for mineral oil, making these new substances attractive as coolants.  Increasing the operational temperature limit for wire insulation is also an attractive upgrade for transformers. Typical wire insulation consists of polyamide. New nanomaterials (aliphatic polyamide) that are reinforced with SiO$_2$, Al$_2$O$_3$, and TiO$_2$ have been patented by Weinberg \cite{Weinberg}. These alternative insulations have better thermal properties than pure polyamide. In addition, carbon nanotube (CNT) materials can potentially replace copper wire, as they have high thermal and electrical conductivity \cite{CONTRERA,raus}. As a method to increase transformer cooling rate, Patel et al.\ have used a magnetic fluid as a coolant. A Mn-Zn ferrite magnetic fluid (TCF-56) can reduce the winding, core, and top oil temperatures by 20$^{\circ}$C, 14$^{\circ}$C, and 21$^{\circ}$C, respectively \cite{PATEL}.

In practice, an Omnimagnet operates for short time intervals, so it is unlikely that a steady-state thermal condition is reached. Thus, calculating the maximum allowable current density (as in Petruska et al.\ \cite{Petruska2014-2}) to produce a desired magnetic field does not provide transient temperature data. In this study, a transient thermal model of an Omnimagnet is developed and validated for the first time. A preliminary version of this work was presented in \cite{confPaper}. The model is applied to determine the relationship between current density and thermal limits. In addition, potential cooling methods are investigated. Finally, a model in state space form is provided to facilitate  control of the magnetic field strength while taking into account transient temperature increase. 

\section*{Thermal lumped capacitance model}

\begin{figure} [t]
\begin{center}
\includegraphics [width=4in] {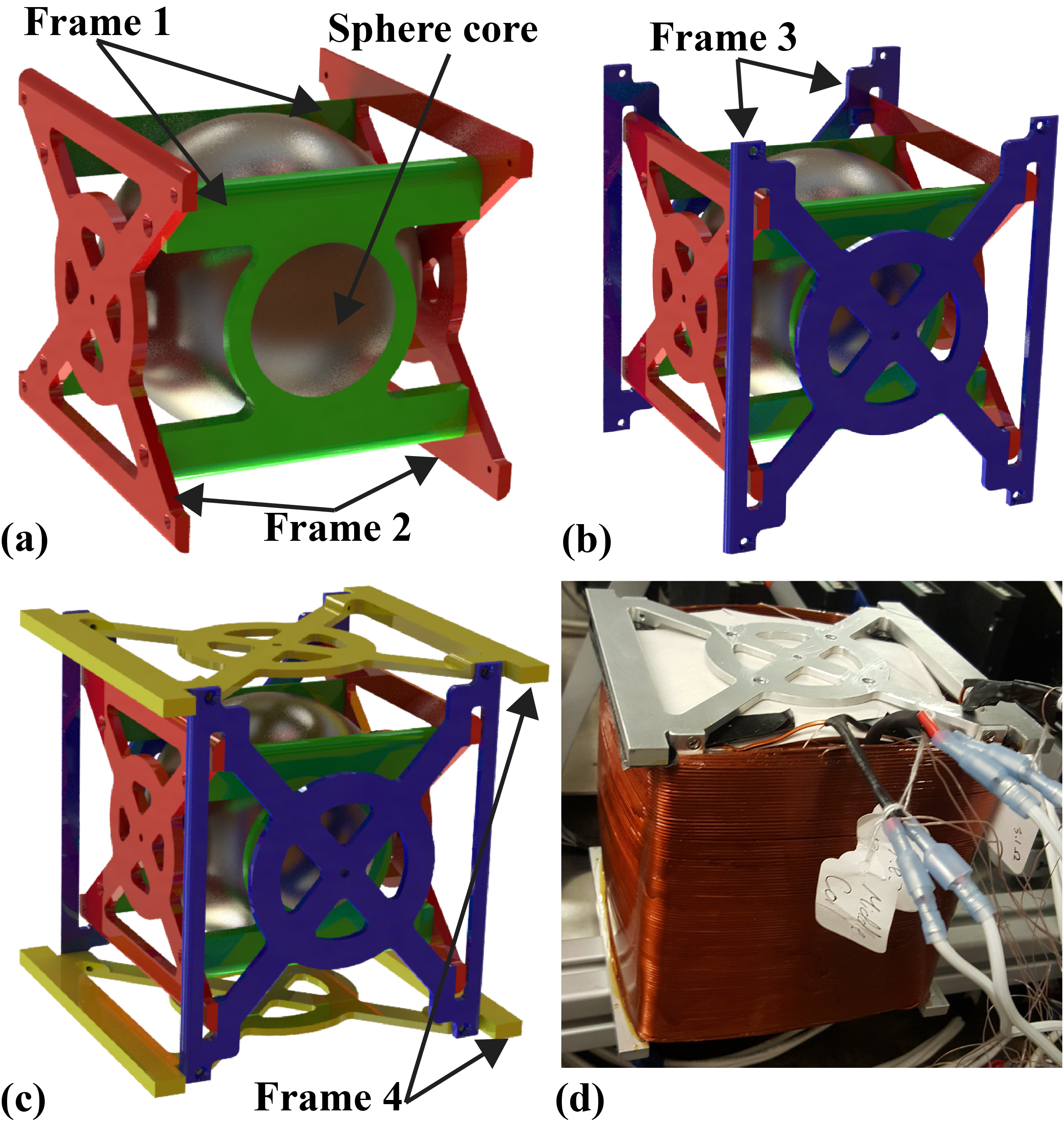}
\end{center}
\caption {OMNIMAGNET COMPONENTS. (a) FRAME 1, FRAME 2, AND THE SPHERICAL CORE; (b) FRAME 3 IS ADDED; (c) FRAME 4 IS ADDED.  THE THREE SOLENDOIDS, WHICH ARE WOUND AROUND THE RESPECTIVE FRAMES, ARE OMITTED FOR CLARITY. THE DIMENSIONS SHOWN ARE FOR ONE PARTICULAR OMNIMAGNET DESIGN, BUT THE DEVICE CAN BE SCALED HOMOTHETICALLY. (d) PHOTO OF AN ASSEMBLED OMNIMAGNET.} 
\label{1}
\end{figure}

The details of an Omnimagnet design are presented in \cite{Petruska2014-2}. The main components include frames, solenoids (coils), wire insulation, and a ferromagnetic core. The ferromagnetic core can be spherical or cubical. An Omnimagnet may be thermally modeled as a set of its elements (see Fig.~\ref{1}).
\begin{itemize}   
\item{$\bullet$}  Frames: Four sets of two identical parallel frames are included (four elements).
\item{$\bullet$}  Solenoids: Three solenoids of copper wire wound around each frame are the main components of an Omnimagnet (three elements).
\item{$\bullet$}  Wire insulation: Solenoid wires are insulated; however, each solenoid is modeled as a copper solid rather than a conglomeration of separate wires. Thus, the wire insulation is also modeled as a single solid surrounding the solenoid conductor (three elements ).
\item{$\bullet$} Ferromagnetic core: A solid ferromagnetic spherical or cubical core is located in the middle of the Omnimagnet to magnify the electromagnetic field; this sphere/cube is considered a separate element (one element). 
\end{itemize}

In addition to the main components, the design may include extra electrical insulation to ensure the safety of the Omnimagnet. These insulation layers are modeled as separate components.

\begin{itemize}  
\item{$\bullet$}  Cover insulation: The outer surface of each solenoid is covered with an additional layer of thin electrically resistive insulation (three elements).
\item{$\bullet$}  Insulating paper or Kapton tape: Thermal insulating paper or Kapton tape is placed between each solenoid and the surrounding frames. 

\item{$\bullet$} Material in the innermost region: The material which fills the deepest part of the Omnimagnet is modeled as an independent element. This gap may be filled with trapped air or any other material. 
\end{itemize}

The four frames, each consisting of two parallel frame sections, and their assembly for a sample Omnimagnet are shown in Fig.~\ref{1}. Inner, middle, and outer solenoids are wound around Frame 1, Frame 2, and Frame 3, respectively. Frame 4 is used as a support to join all the parts together. An image of a complete Omnimagnet is shown in Fig.~\ref{1}(d).  The wire insulation and outer insulating cover insulation are optically transparent; thus, they are not visible in the photo.

Due to the high thermal conductivity of most Omnimagnet components (on the order of 10--100\,W$\cdot$m${^{-1}}$$\cdot$K${^{-1}}$), low convection coefficient (natural convection assumed, 0--10\,W$\cdot$m$^{-2}$$\cdot$K$^{-1}$), and small characteristic length of each component ($< 0.2$\,m), $Bi$ for each element is less than 0.1. Owing to small $Bi$ and symmetry, it is reasonable to model the Omnimagnet using a lumped-capacitance method.  The main assumptions for the application of this method to the elements of the Omnimagnet include: 
\begin{itemize} 
\item {$\bullet$}  Temperature distribution is uniform in each element ($Bi < 0.1$).
\item {$\bullet$}  Conduction between elements is modeled with Fourier's Law, where the differentials are approximated with variable differences.
\item {$\bullet$}  Thermal contact resistance is neglected.
\item {$\bullet$}  The temperature of the surroundings is equal to that of ambient air.
\item {$\bullet$}  All thermal properties of the elements (except the copper coils) are considered to be temperature independent.
\item {$\bullet$}  Radiation heat transfer between different elements is neglected; however, radiation between the surrounding and insulating paper, Frame 4, wire insulation of solenoid 3, cover insulation of the solenoid 3, and outer solenoid (solenoid 3) is included.
\item {$\bullet$} Electrical insulating materials (paper or Kapeton tape) is placed between each solenoid and the surrounding frames. These insulating layers can be incorporated in the model as correction coefficients. 
\item {$\bullet$}  The view factor between solenoid 2 and ambient, as well as the view factor between solenoid 3 and ambient, are both assumed to be unity. 
\end{itemize}

Utilizing these simplifying assumptions and applying an energy balance to each element, $n$ ordinary differential equations (ODEs) are developed for $n$ unknown temperatures. $n$ is the number of elements of the Omnimagnet. The number of elements includes the principal elements (11 elements), the inside trapped material (1 element), and the number of extra electrical insulation layers. The general form of the ODE for each element is:

\setlength{\belowdisplayskip}{0pt} \setlength{\belowdisplayshortskip}{0pt}
\setlength{\abovedisplayskip}{0pt} \setlength{\abovedisplayshortskip}{0pt}
\begin{equation} \label{eq:1}
   \rho_i c{_p}{_i} V_i \frac {dT_i}{dt}=R_iI_{i}^2-\left[{\sum_{j=1,{j\neq i}}^{j= n}}\left(\sum_{m=1}^{m=3} \kappa_jS_{i,j,m}\frac{T_i-T_j}{\Delta x_{j,m}}\right)\right] -\left[{\sum_{j=0|{j=14}}}\left(\sum_{m=1}^{m=3}h_{i,j}S_{i,j,m}(T_i-T_{\mbox{\scriptsize{j}}})\right)\right]- \sum_{m=1}^{m=3}\sigma\epsilon_iS_{i,0,m}(T_i^4-T_{\mbox{\scriptsize{0}}}^4)
\end{equation}

The variables $\rho_i$, $V_i$, $c{_p}{_i}$, $T_i$, $\epsilon_i$, $R_i$ and $I_i$ are density (kg$\cdot$m\textsuperscript{-3}), volume (m\textsuperscript{3}), specific heat capacity at constant pressure (kJ$\cdot$kg\textsuperscript{-1}$\cdot$K\textsuperscript{-1}), temperature (K), emissivity, electrical resistance ($\Omega$), and input current (A) of the \textit{i}\textsuperscript{th} element, respectively. $\kappa_j$ is thermal conductivity (W$\cdot$m\textsuperscript{-1}$\cdot$K\textsuperscript{-1}) of the \textit{j}\textsuperscript{th} element. \textit{I}\textsubscript{\textit{i}} is nonzero only for the three solenoids. Time (s) is represented by \textit{t}.  \textit{S}\textsubscript{\textit{i,j,m}} and $\Delta$\textit{x}\textsubscript{\textit{j,m}} are contact surface area (m\textsuperscript{2}) and thickness (m) between the \textit{i}\textsuperscript{th} and \textit{j}\textsuperscript{th} elements  in the \textit{m} direction (\textit{x} = 1, \textit{y} = 2, or \textit{z} = 3), respectively. Note that \textit{S}\textsubscript{\textit{i,j,m}} = \textit{S}\textsubscript{\textit{j,i,m}}. $h_{i,j}$ is the convection coefficient (W$\cdot$m\textsuperscript{-2}$\cdot$K\textsuperscript{-1}) between the \textit{i}\textsuperscript{th} element and the ambient ($j$ = 0) or the inner air ($j$ = 14). $h_{i,j}$ is a function of temperature and the orientation of the surface. Based on the orientation of the heated surface $h_{i,j}$ can be estimated by the convection coefficient correlation in an enclosure $(h_{i,j})_{encl}$, over a vertical surface $(h_{i,j})_v$, above a horizontal surface $(h_{i,j})_{ha}$, or below a horizontal surface $(h_{i,j})_{hb}$. All heat transfer coefficient correlations are provided in Appendix A for a specific Omnimagnet. 

$T_{\mbox{\scriptsize{14}}}$ is the temperature of the innermost trapped fluid (K). Note that if the gap inside the inner solenoid (Solenoid 1) and the middle solenoid (Solenoid 2) is filled with a solid, this term is replaced by a conduction term.  $\sigma$ is the Stefan-Boltzmann constant ($5.67\times10^{-8}$\,W$\cdot$m\textsuperscript{-2}$\cdot$K\textsuperscript{-4}) and $T_{\mbox{\scriptsize{0}}}$ is ambient temperature (K) of the air surrounding the device.~The term on the left-hand side of Eq.~(\ref{1}) is the storage term for the \textit{i}\textsuperscript{th} element.  The first term on the right-hand side of Eq.~(\ref{1}) represents heat source. Source terms are applicable only to the equations related to the three solenoids. The second term represents conduction between elements, the third term represents convection to the inside trapped fluid, or the convection to ambient if the component is exposed to the ambient air, and the fourth term represents radiation exchange with the surroundings. Radiation is included in the equations related to the outer solenoid, the outermost frame (Frame 4), and the insulating paper.

Equation ~(\ref{1}) is applied to each of the $n$ elements to produce $n$ ODEs with $n$ unknown element temperatures. The model requires input data such as initial temperature, volume and thickness of each element, contact surface area between two adjacent elements, thermal properties, convection coefficients, and electrical resistance and input current of each solenoid. An example of the thermal model applied to a specific Omnimagnet is presented in Appendix A. The equation set is solved using MATLAB (function ode45). The ode45 function uses the last time step to provide the results for the next time step using an explicit Runge-Kutta (4,5) method \cite{Shampine}. A state space format of the equations is provided in Appendix B. Simulations from the model should be validated and assessed for accuracy by comparisons to experimental data. An experimental apparatus and test procedure for model validation are presented in the following section.

\section*{Experimental setup}

\begin{figure}[h]
\begin{center}
\includegraphics  [width=3.8in] {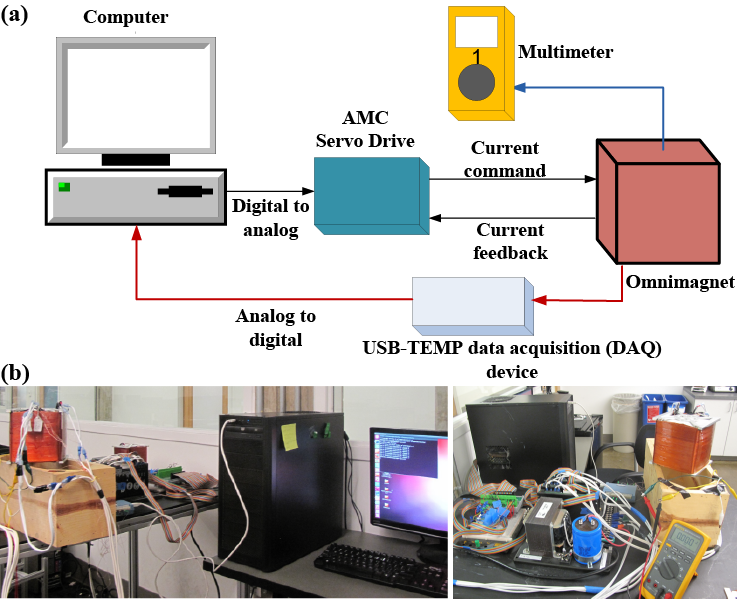}
\end{center}
\caption {EXPERIMENTAL SETUP.  (a) SCHEMATIC, (b) PHOTOS.}
\label{2}
\end{figure}

A schematic of the experimental setup is depicted in Fig.~\ref{2}. An AMC high-frequency PWM servo drive provides an effecively constant electrical input current to the solenoids. Two Omega type-K thermocouples are implanted on opposite sides of each solenoid, near the coil center, to measure temperature. The location of each thermocouple depends on the number of windings in each solenoid. The thermocouples are implanted at the middle layer if there are an even number of layers, and one wire towards the core from that if there is an odd number of layers. Two similar thermocouples are used to measure the ambient air temperature and the air temperature beneath the center of the Omnimagnet. Electrical resistance of each solenoid is measured using a Fluke 87 True-RMS Multimeter (accuracy $\pm$  2\,$\Omega$) at the beginning of each experiment. All temperature data are recorded every 0.5\,s and the electrical potential difference of each solenoid is measured every 5 minutes.

The computer and AMC servo drive together control the input current and maintain it at the desired value to each solenoid.~As temperature increases, the copper wire resistance changes, requiring modifications to the applied voltage to achieve a constant current.~The control loop functions between the computer and the solenoids and does not require temperature feedback. In Fig.~\ref{2}, the dashed line between the data acquisition system (DAQ) and the computer indicates temperature data acquired and stored on the computer. A single thermocouple from the same batch of identical thermocouples was calibrated using a two-point method. 

The lowest melting point in the entire Omnimagnet system is selected as the critical temperature. The experiment is stopped when the maximum temperature is 5$^{\circ}$C lower than the critical temperature or after 2~hr. The maximum time that the Omnimagnet is powered depends on the input current; higher input current leads to shorter safe operating time. Experimental data from this process are used to validate simulation results, as presented in Appendix A.

To validate the model, 40 independent experiments were conducted on the lab-scale Omnimagnet. In each experiment, temperatures of the three solenoids were measured at two separate locations (six measurements total) as a function of time. Measurements were repeated for three current levels and different combinations of powered solenoids (20 heating experiments).~Passive Omnimagnet cooling was also considered (20 experiments).~The standard deviation of the mean of the thermocouple temperatures was determined to be 0.03$^{\circ}$C. 
\section*{Validation and optimization}
Omnimagnets are manually wound; therefore, some uncertainty exists in values required as input data to the model. For instance, the exact value of the contact surface area between different elements, the thermal contact resistance, and the thickness of the various insulation layers are not easily measurable and are not reported.  Given these unknown parameters, the following assumptions are made to the model initially:
\begin{itemize} 
\item {$\bullet$} Contact resistance between different elements is neglected.
\item {$\bullet$} Each solenoid is in perfect contact with its frame and with the adjacent solenoid(s). Thus, the maximum contact area between elements is assumed.
\item {$\bullet$} Thickness of all elements is assumed to be constant. Shrinkage or expansion due to winding pressure or temperature increase is neglected.
\end{itemize}
The model is solved applying these assumptions and the resulting temperature data are compared with experimental data. Conditions in the model are set to match those used in the experiments. If the differences in solenoid temperature between the experiment and simulations are acceptable, this model can be applied for further studies or for controlling the system. But, due to the mentioned uncertainties, better performance can be achieved by modifying the model using experimental data.~To reduce the error, and to ultimately create a more accurate model, a set of coefficients is added to the basic code. These coefficients are applied to the terms with uncertain dimensions (e.g., thickness, contact area) and where there is a high probability that contact resistance may be present. In addition, some coefficients are used as correction factors in the convection heat transfer terms to better estimate convection coefficients. All assumptions and simplifications presented in the previous paragraph are incorporated through these correction coefficients.

The number of correction coefficients and the appropriate terms to augment in the model depends on the structure of any specific Omnimagnet. As an initial step, a correction coefficient is added to each term and an optimization process is performed. If a correction coefficient is found to be small after optimization, that coefficient can be neglected. Only the correction coefficients that cause a significant change to the final results (minimization of the difference between model and experimental data) in the optimization process are incorporated into the model. 

\begin{figure}[t]
\begin{center}
\includegraphics [width=4in]{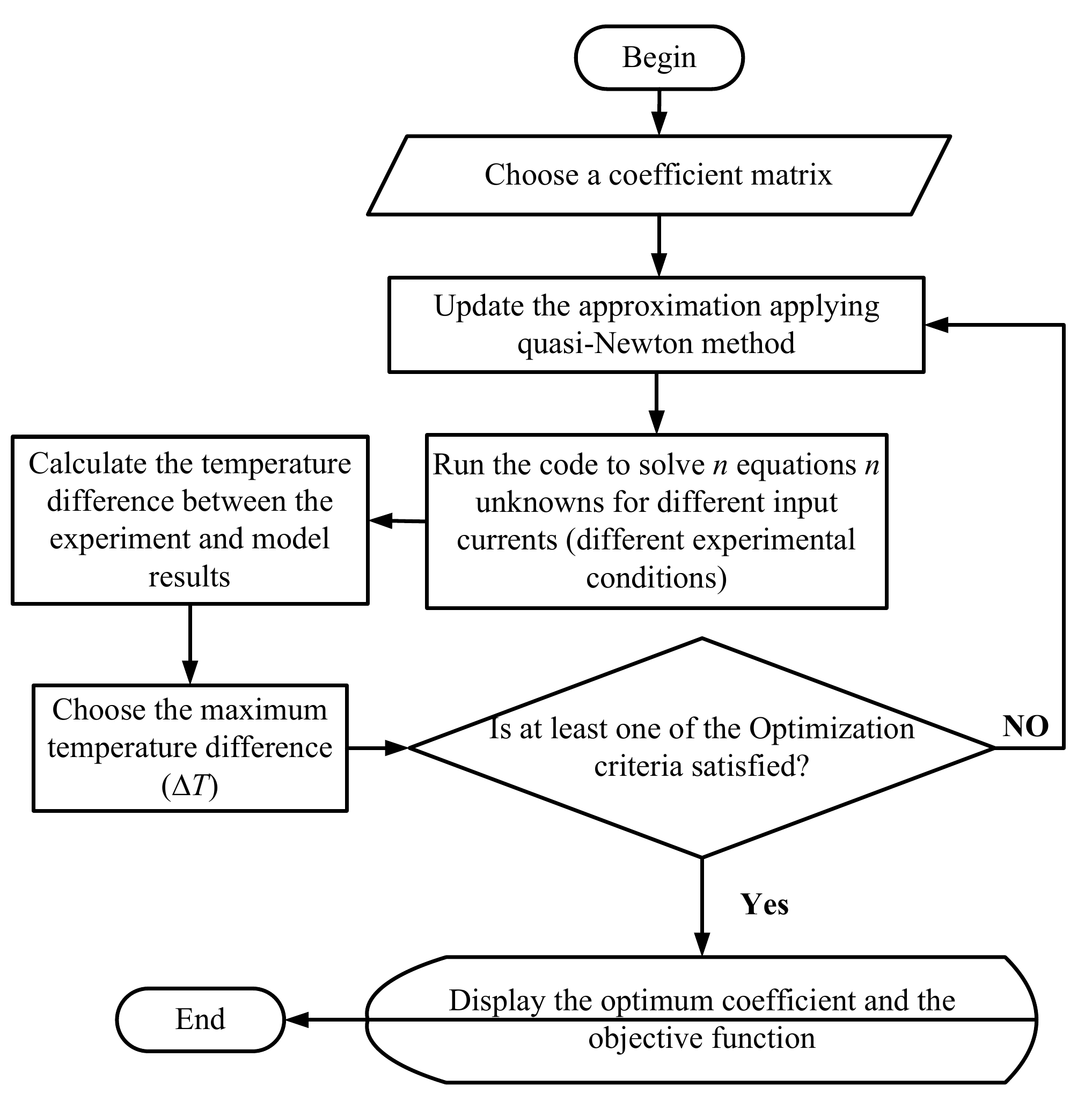}
\end{center}
\caption {OPTIMIZATION ALGORITHM TO DETERMINE MODIFICATION COEFFICIENTS.}
\label{3}
\end{figure}

As noted, an optimization process is performed to determine appropriate values for the correction coefficients based on minimization of the solenoid temperature differences between the experiments and simulations. During the optimization process, transient solenoid temperatures for seven experiments corresponding to the seven combinations of powered solenoids, are evaluated and compared to simulation data determined under similar conditions.~A MATLAB optimization toolbox (using the fmincon `interior point algorithm' \cite{kiusalaas_2015}) is used to minimize the maximum solenoid temperature difference by varying the coefficients using a quasi-Newton method. The algorithm is illustrated in Fig.~\ref{3}. The final result is a semi-empirical model, based on the lumped capacitance method, for the Omnimagnet. The coefficients for the conduction terms can vary between 0 to 1. On the other hand, the coefficients for the convection terms can be larger than one, but the upper limit for these coefficients should not exceed the upper limit for the type of the convection used to cool the Omnimagnet (e.g., for natural convection in air the convection coefficient should not exceed 25 W$\cdot$m\textsuperscript{-2}$\cdot$K\textsuperscript{-1} \cite{Incropera}. 

\begin{figure}[h!]
\begin{center}
\includegraphics  [width=\linewidth] {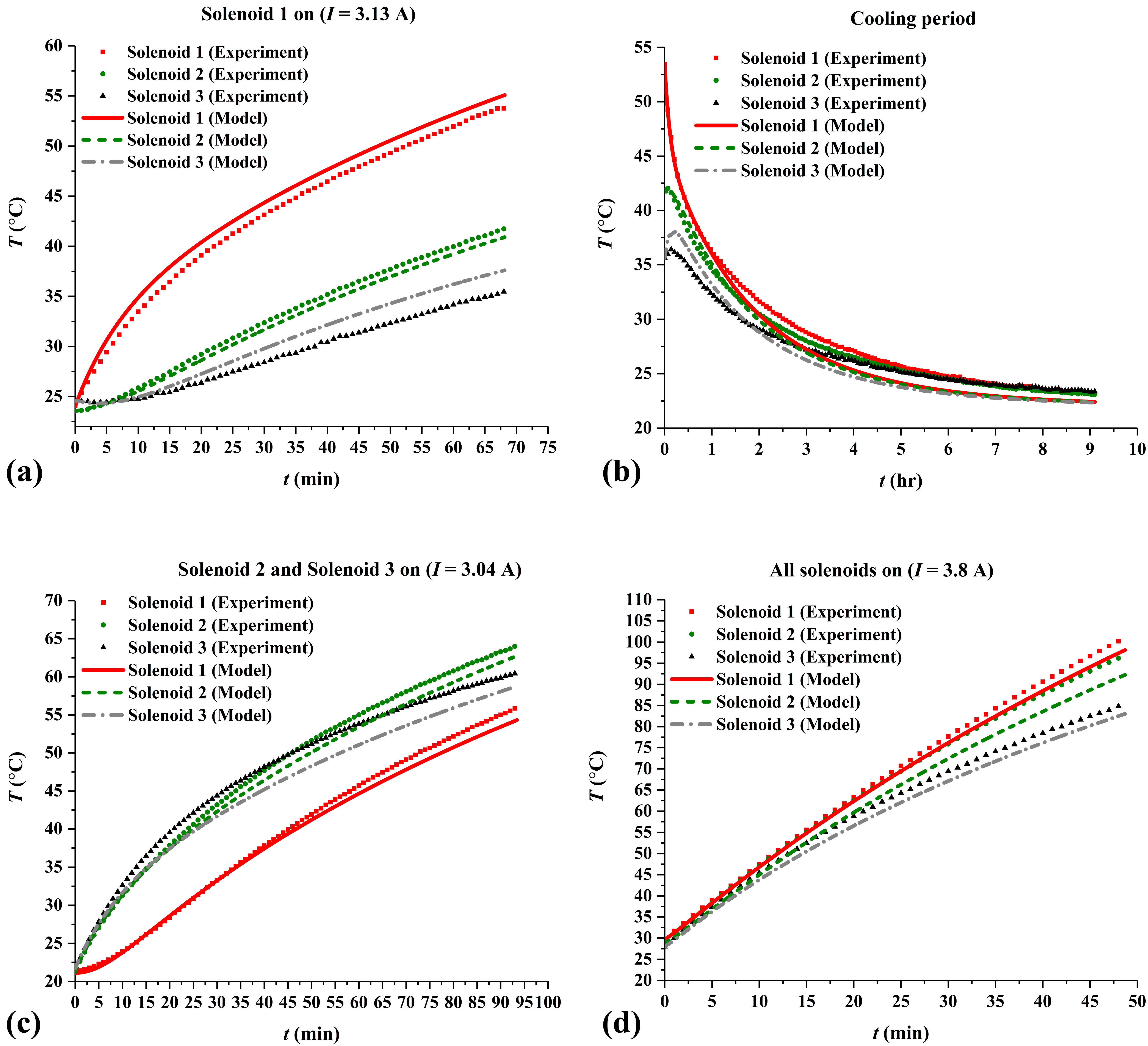}
\end{center}
\caption{EXPERIMENTAL VALIDATION OF THE MODEL. (a) INNER SOLENOID ON; (b) COOLING PERIOD, NO CURRENT FLOW; (c) MIDDLE AND OUTER SOLENOIDS ON; (d) ALL THREE SOLENOIDS ON.}
 \label{fig4}
\end{figure}

The model is applied to the Omnimagnet used in the experiments. Details of the input data for the Omnimagnet are presented in Appendix A. Following the optimization process, the maximum solenoid temperature difference between the simulations and experiments is less than 4$^{\circ}$C at temperatures less than 120$^{\circ}$C. Comparisons of simulation and experiment transient temperatures for the three solenoids are shown in Fig.~\ref{fig4}. In Fig.~\ref{fig4}(a), only the inner solenoid (Solenoid 1) is powered (at 3.13\,A). The maximum root-mean-square error (RMSE) and normalized root-mean-square error (NRMSE) in this case are 1.7$^{\circ}$C and 4.7\%, respectively. During cooling, no electrical current is flowing in any part of the system (Fig.~\ref{fig4}(b)). The model accurately predicts solenoid temperatures under conditions of natural convection and radiation cooling (RMSE = 1.2$^{\circ}$C, NRMSE = 3.1\%).  Transient solenoid temperature response when current is applied to the middle (Solenoid 2) and outer solenoids (Solenoid 3) (3.04\,A in each) is shown in Fig.~\ref{fig4}(c).  Here the error (RMSE = 0.43$^{\circ}$C, NRMSE = 1\%) is less than the case when only Solenoid 1 is powered (Fig.~\ref{fig4}(a)).  Finally, the transient solenoid temperature response when all three solenoids are powered at 3.8\,A is presented in Fig.~\ref{fig4}(d), where the maximum RMSE and NRMSE are 0.6$^{\circ}$C and 0.6\%, respectively. These errors are deemed acceptable given the uncertainties in some of the input data and the assumptions inherent to the lumped capacitance method. When using an Omnimagnet in practice, a factor of safety can be included that accounts for this level of modeling error. 


\section*{Results and discussion}
\begin{figure*}[t]
\begin{center}
\includegraphics  [width=4in] {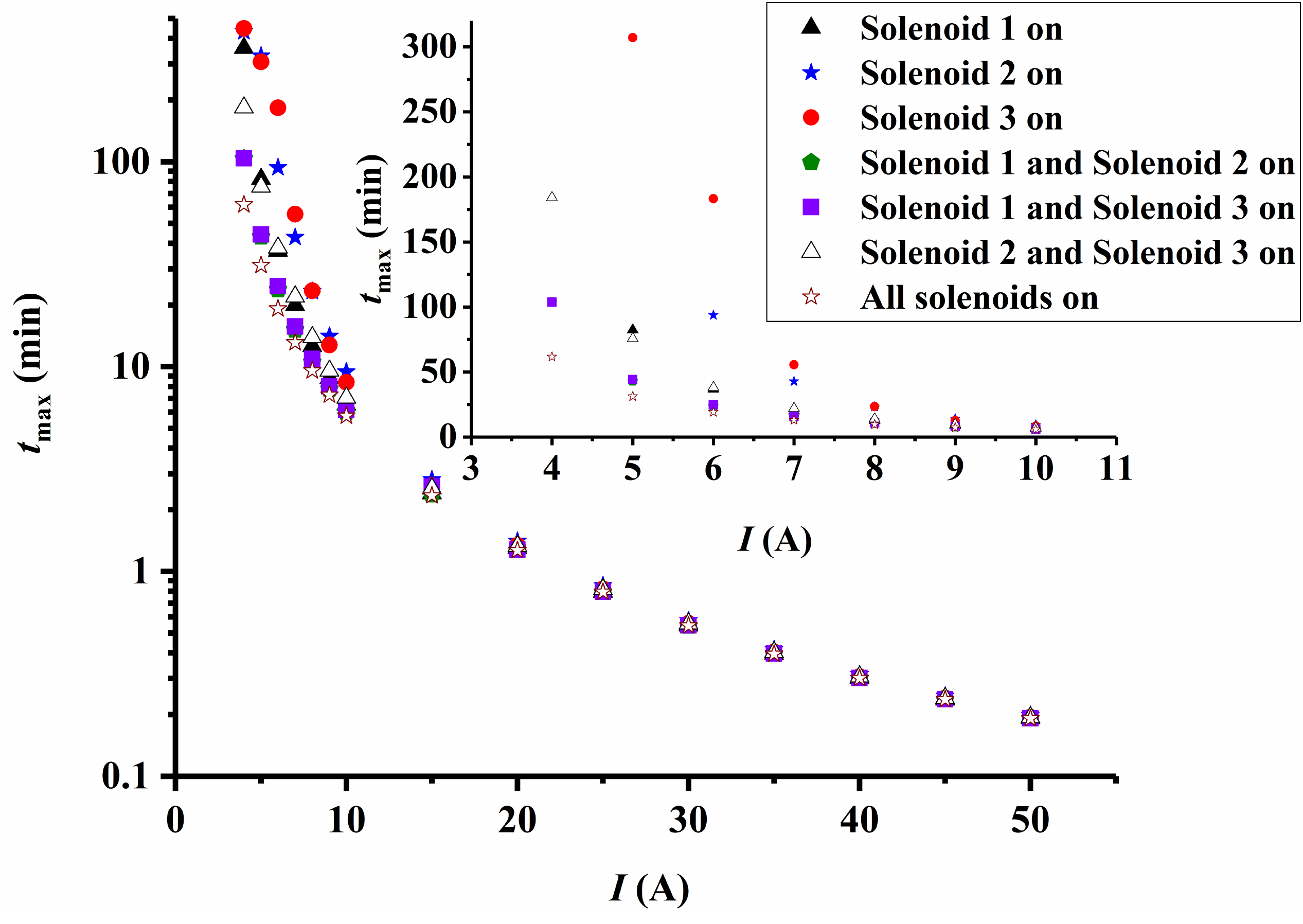}
\end{center}
\caption {MAXIMUM OPERATIONAL TIME FOR THE OMNIMAGNET UNDER SEVEN DIFFERENT COMBINATIONS OF POWERED SOLENOIDS IN A RANGE OF CURRENT FROM 0 TO 50\,A AND NATURAL-CONVECTION COOLING.}
\label{5}
\end{figure*}

After validation, the model is used to study the thermal behavior of the Omnimagnet under different conditions. Of most interest is the maximum time $t_{\mbox{\scriptsize{max}}}$ the Omnimagnet can be powered before the maximum temperature $T_{\mbox{\scriptsize{max}}}$ in the device reaches the temperature limit (glass temperature) where the outer insulation would begin melting, which for the Omnimagnet prototype (see Appendix A) is 115$^{\circ}$C.  $t_{\mbox{\scriptsize{max}}}$ data are shown in Fig.~\ref{5} for seven cases corresponding to seven combinations of solenoids carrying current $I$. 

As shown in Fig.~\ref{5}, when cooling is achieved by natural convection and radiation with the surroundings, the worst-case scenario occurs when all three solenoids are powered simultaneously.  The inner solenoid heats more quickly than the other two, due to the high thermal resistance between the inner solenoid (solenoid 1) and the ambient. According to the data in Fig.~\ref{5}, free convection is not very effective in removing excess heat from the Omnimagnet, which limits the time the device can be powered.  Thus, more effective cooling mechanisms should be considered. Increasing the convection heat transfer coefficient on the surface of the middle and outer coils, which are exposed to the ambient, is an option to decrease $T_{\mbox{\scriptsize{max}}}$ and subsequently increase $t_{\mbox{\scriptsize{max}}}$.  The effect of the convection heat transfer coefficient is discussed in the next section.

\hfill
\subsection*{Effect of \textbf{\textit{h}} for heating and cooling}
\begin{figure}[t]
\begin{center}
\includegraphics  [width=3in] {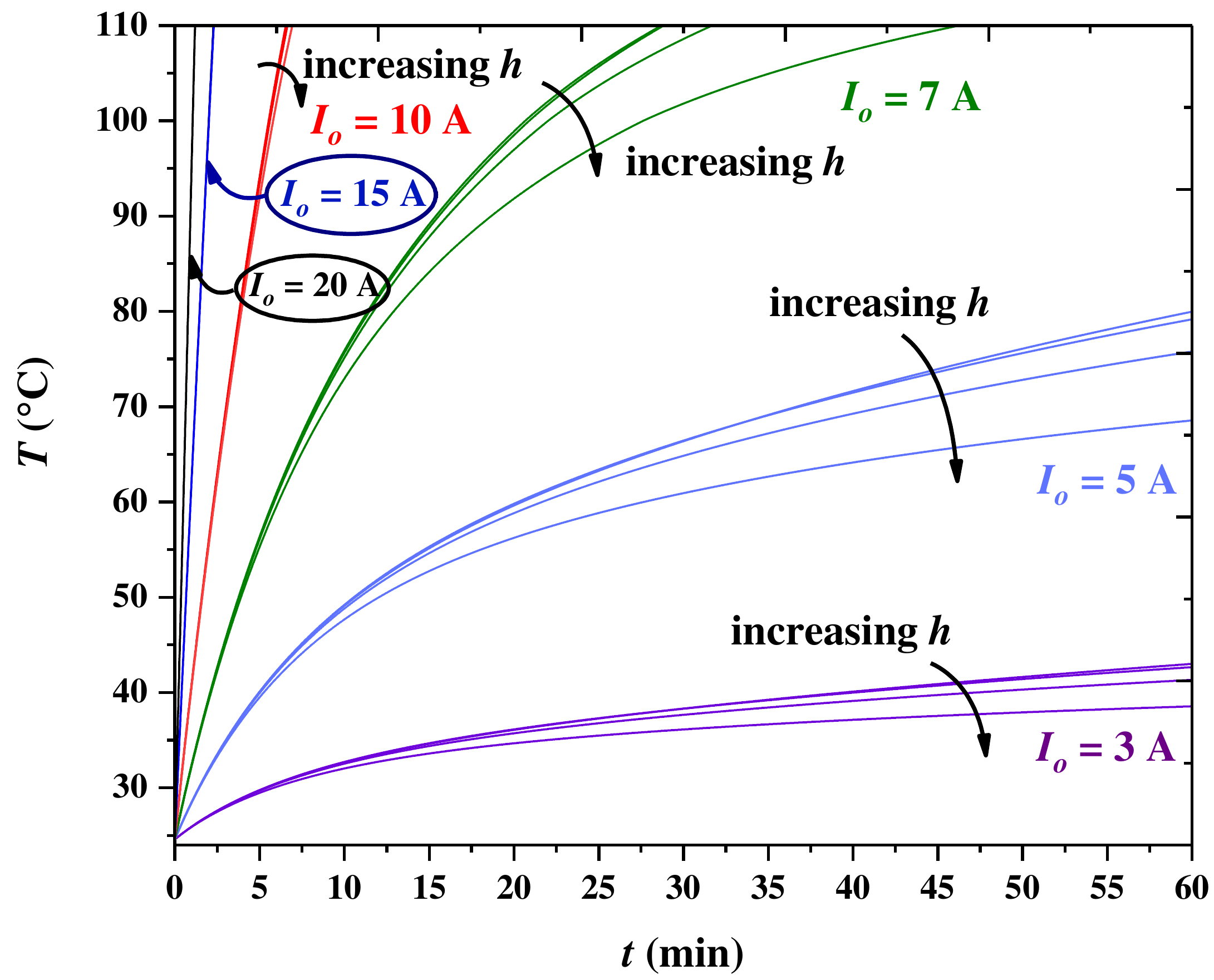}
\end{center}
\caption {THE EFFECT OF HEAT TRANSFER COEFFICIENT ON THE OUTER SOLENOID TRANSIENT TEMPERATURE (ONLY SOLENOID ON) AT SIX DIFFERENT INPUT CURRENTS (3, 5, 7, 10, 15, AND 20\,A).  ARROWS INDICATE INCREASING $h$ (5, 250, 500, AND 1000 W$\cdot$m$^{-2}$$\cdot$K$^{-1}$ )}
\label{6}
\end{figure}

The inner solenoid is not exposed to the ambient so it is not directly affected by convective cooling from the outer surfaces of the Omnimagnet. On the other hand, all four external surfaces of the outer solenoid and two of the four external surfaces of the middle solenoid are exposed to convective cooling with ambient air. Therefore, it is expected that increasing the convective heat transfer coefficient will directly affect the transient temperature response of the middle and outer solenoids while the inner solenoid will only be indirectly influenced. The effect of $h$ on the temperature of the solenoids is studied in both cooling and heating modes. A set of four different values of $h$ are selected for the study.  A heat transfer coefficient equal to 5 Wm$^{-2}$K$^{-1}$ is chosen as a representation of natural convection in air. A value of 250 Wm$^{-2}$K$^{-1}$ represents forced convection in air, and 500 and 1000 Wm$^{-2}$K$^{-1}$ correspond to natural convection in a liquid media, which can be produced by nanofluids as mentioned in the Introduction. The effect of $h$ for six different values of input current applied only to the outer solenoid is shown in Fig.~\ref{6}. Generally, the heating rate for the solenoid with the highest temperature (Solenoid 3) is slowed as $h$ increases, but this effect is tempered at higher input current. For large I (depending on size and structure of the Omnimagnet) this effect is negligible. At lower current, the input power is of the same order of magnitude as the convection heat transfer rate, which enables control of the Omnimagnet temperature. According to Eq.~(\ref{1}), doubling the current leads to four times the power. As a result, for larger values of $I$, increases in convection coefficient have less effect on reducing the rate of temperature increase, making a maximum safe operating temperature impossible.  It should also be noted that for relatively high input current (it depending on size and structure of the Omnimagnet), Solenoid 3 heats very rapidly and the safe operating temperature is achieved quickly.  For these power levels, heat dissipation by convective cooling is overwhelmed by the input power and changes in $h$ are inconsequential. It can be concluded that for high $I$, forced convection from the outer surfaces of the Omnimagnet by itself is not a suitable method to slow the heating process and to prevent undesirable effects of high temperature.

\begin{figure}[t]
\begin{center}
\includegraphics  [width=3in] {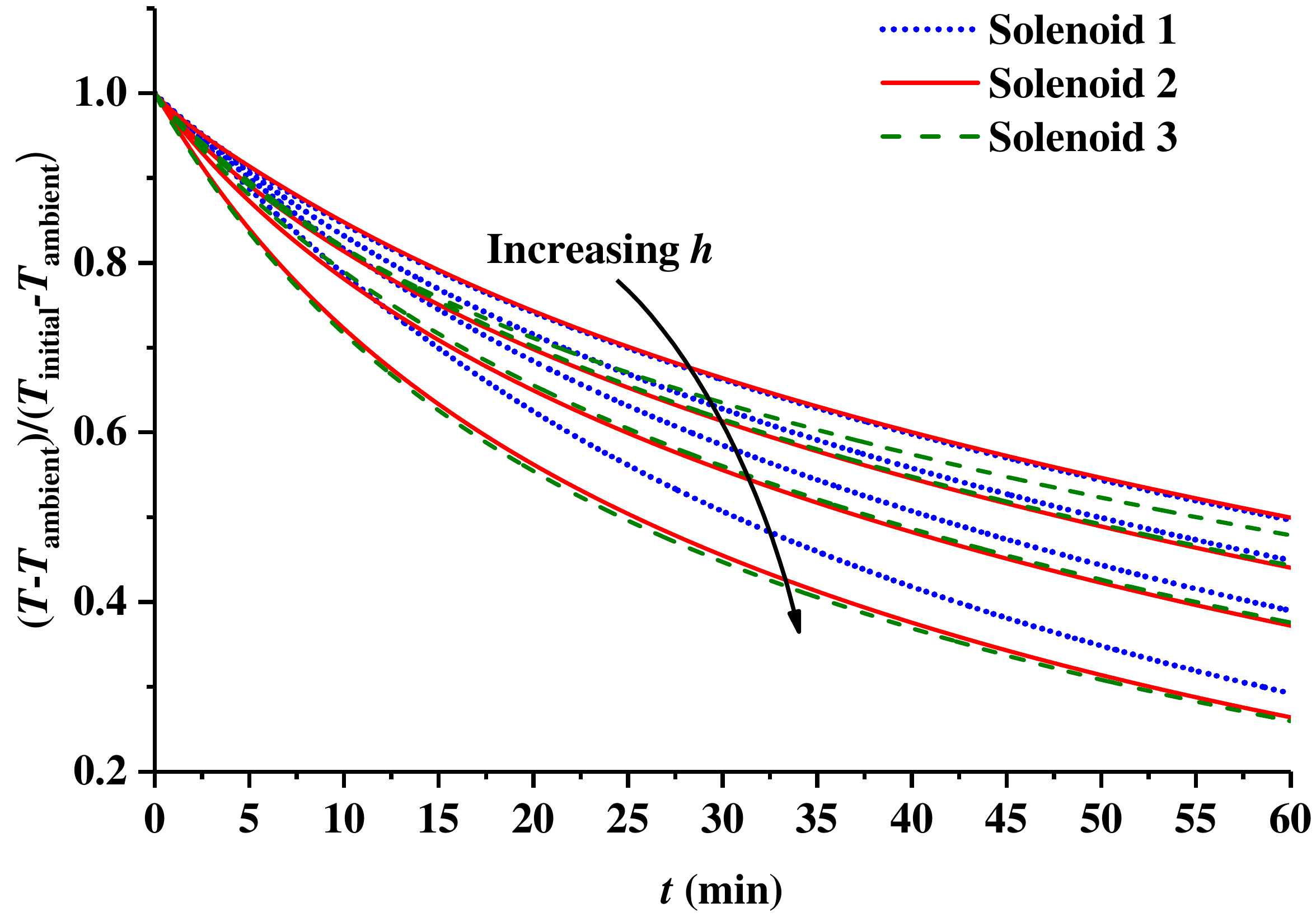}
\end{center}
\caption {THE EFFECT OF HEAT TRANSFER COEFFICIENT ON THE TEMPERATURE OF THE THREE SOLENOIDS WITH NO INPUT CURRENT.  THE ARROW INDICATES INCREASING $h$ (5, 250, 500, AND 1000 W$\cdot$m$^{-2}$$\cdot$K$^{-1}$).}
\label{7}
\end{figure}

While convective cooling of the Omnimagnet is found to be insufficient for extending the operational time at relatively high input power, convective cooling should be effective when none of the solenoids is powered and the objective is to cool the entire device to an acceptable temperature level.  To study the validity of this hypothesis, the transient response of non-dimensional temperature 
of the three solenoids is considered with no power applied to the device (Fig.~\ref{7}). To calculate the initial temperature, it is assumed that all solenoids are powered for 60 min with $I$ = 3\,A. The temperature at the last time step is used as the initial temperature of the cooling period. Initial temperatures for Solenoids 1, 2, and 3 are 75.3$^{\circ}$C, 71.1$^{\circ}$C, and 66.0$^{\circ}$C, respectively. A non-dimensional temperature is chosen to display the temporal response as the trends are self-similar for different initial temperatures.  As expected, cooling rates are greater for increasing values of $h$.  This is especially true for forced convection with air or a liquid. As expected, the outer solenoid (Solenoid 3) with its entire outer surface exposed to the convective environment cools at the highest rate.  The data in Fig.~\ref{7} indicate that the cooling rate on the inner solenoid (Solenoid 1) is lower than the other two. The inner solenoid does not have direct contact with the ambient air, while the middle solenoid (Solenoid 2) has less surface area exposed to convection in comparison to the outer solenoid (Solenoid 3). While the cooling process is reasonably effective, it would be advantageous to find other means to reduce $T_{\mbox{\scriptsize{max}}}$, increase $t_{\mbox{\scriptsize{max}}}$, and decrease the cooling time to achieve a desired temperature level.  

\section*{Conclusion}
For the first time, thermal performance of an Omnimagnet is studied using a lumped capacitance model, which is validated with experimental data from a single Omnimagnet.~An optimized model, which includes correction coefficients for the Omnimagnet used in the validation, is found to be in good agreement with the experimental data.  The maximum root mean square error of the model is approximately 8$\%$ (4$^{\circ}$C). Based on a parametric study that considers all seven possible combinations of powered solenoids, the safe range of input current, where the maximum temperature of the system does not reach the minimum melting point of the Omnimagnet components, is calculated. Increases to the external heat transfer coefficient are found to be beneficial only at low input current.  With no power applied to the device, external convection is found to be much more beneficial for heat dissipation although cooling time is still significant. 
Given the ineffectiveness of external convection, active cooling of device components, such as the frame or the interior sphere, should be considered to improve overall thermal performance of the Omnimagnet. The thermal model is linearized and presented as a state-space equation that can be applied in a Kalman filter controller. This model may be used as a basic thermal management tool for any type of omnidirectional electromagnetic device.

\section*{Acknowledgment}
Research reported in this publication was supported by the National Institute on Deafness and Other Communication Disorders of the National Institutes of Health under Award Number R01DC013168. The content is solely the responsibility of the authors and does not necessarily represent the official views of the National Institutes of Health. We would like express our gratitude to Dr.\ Lisandro Leon for his assistance with the experiments. We also acknowledge the Center for High Performance Computing at the University of Utah for their support and resources, and in particular thank Dr.\ Martin Cuma for his assistance parallelizing the optimization code. 

\renewcommand*\labelenumi{[\theenumi]}

\bibliographystyle{asmems4}
\bibliography{asme2e}

\begin{thebibliography}{10}

\bibitem{Petruska2014-2}
Petruska, A.~J., and Abbott, J.~J., 2014.
\newblock ``Omnimagnet: An omnidirectional electromagnet for controlled
  dipole-field generation''.
\newblock {\em IEEE Transactions on Magnetics, {\bf 50}}(7), pp.~1--10.

\bibitem{Petruska2014}
Petruska, A.~J., Mahoney, A.~W., and Abbott, J.~J., 2014.
\newblock ``Remote manipulation with a stationary computer-controlled magnetic
  dipole source''.
\newblock {\em IEEE Transactions on Robotics, {\bf 30}}(5), October,
  pp.~1222--1227.

\bibitem{Clark}
Clark, J.~R., Leon, L., Warren, F.~M., and Abbott, J.~J., 2012.
\newblock ``Magnetic guidance of cochlear implants: Proof-of-concept and
  initial feasibility study''.
\newblock {\em ASME Journal of Medical Devices, {\bf 6}}(3), August,
  pp.~035002--035002--8.

\bibitem{Leon1}
Leon, L., Warren, F.~M., and Abbott, J.~J., 2018.
\newblock ``An in-vitro insertion-force study of magnetically guided
  lateral-wall cochlear-implant electrode arrays''.
\newblock {\em Otology and Neurotology, {\bf 39}}(2), February, pp.~e63--e73.

\bibitem{Leon2}
Leon, L., Warren, F.~M., and Abbott, J.~J., 2018.
\newblock ``Optimizing the magnetic dipole-field source for magnetically guided
  cochlear-implant electrode-array insertions''.
\newblock {\em Journal of Medical Robotics Research, {\bf 3}}(1), January,
  pp.~1--37.

\bibitem{campelo1}
{Campelo}, H. M.~R., {Quintela}, M.~A., {Torriano}, F., {Labbé}, P., and
  {Picher}, P., 2016.
\newblock ``Numerical thermofluid analysis of a power transformer disc-type
  winding''.
\newblock In 2016 IEEE Electrical Insulation Conference (EIC), pp.~362--365.

\bibitem{GASTELURRUTI}
Gastelurrutia, J., Ramos, J.~C., Larraona, G.~S., Rivas, A., Izagirre, J., and
  del Río, L., 2011.
\newblock ``Numerical modelling of natural convection of oil inside
  distribution transformers''.
\newblock {\em Applied Thermal Engineering, {\bf 31}}(4), March, pp.~493 --
  505.

\bibitem{HOSSEINI}
Hosseini, R., Nourolahi, M., and Gharehpetian, G., 2008.
\newblock ``Determination of od cooling system parameters based on thermal
  modeling of power transformer winding''.
\newblock {\em Simulation Modelling Practice and Theory, {\bf 16}}(6), July,
  pp.~585 -- 596.

\bibitem{MUFUTA}
Mufuta, J.~M., and van~den Bulck, E., 2000.
\newblock ``Modelling of the mixed convection in the windings of a disc-type
  power transformer''.
\newblock {\em Applied Thermal Engineering, {\bf 20}}(5), April, pp.~417 --
  437.

\bibitem{TSILI}
Tsili, M.~A., Amoiralis, E.~I., Kladas, A.~G., and Souflaris, A.~T., 2012.
\newblock ``Power transformer thermal analysis by using an advanced coupled 3d
  heat transfer and fluid flow fem model''.
\newblock {\em International Journal of Thermal Sciences, {\bf 53}}, March,
  pp.~188 -- 201.

\bibitem{ALLAHBAKHSHI}
Allahbakhshi, M., and Akbari, M., 2016.
\newblock ``Heat analysis of the power transformer bushings using the finite
  element method''.
\newblock {\em Applied Thermal Engineering, {\bf 100}}, May, pp.~714 -- 720.

\bibitem{SMOLKA}
Smolka, J., 2013.
\newblock ``{CFD}-based 3-{D} optimization of the mutual coil configuration for
  the effective cooling of an electrical transformer''.
\newblock {\em Applied Thermal Engineering, {\bf 50}}(1), January, pp.~124 --
  133.

\bibitem{BEIZA}
Beiza, M., Ramos, J.~C., Rivas, A., Antón, R., Larraona, G.~S., Gastelurrutia,
  J., and de~Miguel, I., 2014.
\newblock ``Zonal thermal model of the ventilation of underground transformer
  substations: Development and parametric study''.
\newblock {\em Applied Thermal Engineering, {\bf 62}}(1), January, pp.~215 --
  228.

\bibitem{campelo2}
Campelo, H., Lopez-Fernandez, X.~M., Picher, P., and Torriano, F., 2013.
\newblock ``Advanced thermal modelling techniques in power transformers. review
  and case studies.''.
\newblock In Conference: Advanced Research Workshop on Transformers, pp.~6--21.

\bibitem{CODDE}
Coddé, J., der Veken, W.~V., and Baelmans, M., 2015.
\newblock ``Assessment of a hydraulic network model for zig–zag cooled power
  transformer windings''.
\newblock {\em Applied Thermal Engineering, {\bf 80}}, April, pp.~220 -- 228.

\bibitem{RODRIGUEZ}
Rodriguez, G.~R., Garelli, L., Storti, M., Granata, D., Amadei, M., and
  Rossetti, M., 2017.
\newblock ``Numerical and experimental thermo-fluid dynamic analysis of a power
  transformer working in onan mode''.
\newblock {\em Applied Thermal Engineering, {\bf 112}}, February, pp.~1271 --
  1280.

\bibitem{TORRIANO}
Torriano, F., Chaaban, M., and Picher, P., 2010.
\newblock ``Numerical study of parameters affecting the temperature
  distribution in a disc-type transformer winding''.
\newblock {\em Applied Thermal Engineering, {\bf 30}}(14), October, pp.~2034 --
  2044.

\bibitem{Lecuna}
{Lecuna}, R., {Delgado}, F., {Ortiz}, A., {Castro}, P.~B., {Fernandez}, I., and
  {Renedo}, C.~J., 2015.
\newblock ``Thermal-fluid characterization of alternative liquids of power
  transformers: A numerical approach''.
\newblock {\em IEEE Transactions on Dielectrics and Electrical Insulation, {\bf
  22}}(5), October, pp.~2522--2529.

\bibitem{Liu}
{Liu}, C., {Ruan}, J., {Wen}, W., {Gong}, R., and {Liao}, C., 2016.
\newblock ``Temperature rise of a dry-type transformer with quasi-3d
  coupled-field method''.
\newblock {\em IET Electric Power Applications, {\bf 10}}(7), July,
  pp.~598--603.

\bibitem{CONTRERA}
Contreras, J., Rodriguez, E., and Taha-Tijerina, J., 2017.
\newblock ``Nanotechnology applications for electrical transformers—a
  review''.
\newblock {\em Electric Power Systems Research, {\bf 143}}, February, pp.~573
  -- 584.

\bibitem{Fontes}
Fontes, D.~H., Ribatski, G., and Filho, E. P.~B., 2015.
\newblock ``Experimental evaluation of thermal conductivity, viscosity and
  breakdown voltage ac of nanofluids of carbon nanotubes and diamond in
  transformer oil''.
\newblock {\em Diamond and Related Materials, {\bf 58}}, September, pp.~115 --
  121.

\bibitem{Guan}
{Guan}, W., {Jin}, M., {Fan}, Y., {Chen}, J., {Xin}, P., {Li}, Y., {Dai}, K.,
  {Zhang}, H., {Huang}, T., and {Ruan}, J., 2014.
\newblock ``Finite element modeling of heat transfer in a nanofluid filled
  transformer''.
\newblock {\em IEEE Transactions on Magnetics, {\bf 50}}(2), February,
  pp.~253--256.

\bibitem{Modestov}
Modestov, M., Kolemen, E., Fisher, A., and Hvasta, M., 2017.
\newblock ``Electromagnetic control of heat transport within a rectangular
  channel filled with flowing liquid metal''.
\newblock {\em Nuclear Fusion, {\bf 58}}(1), November, pp.~016009--016009--9.

\bibitem{PATEL}
Patel, J., Parekh, K., and Upadhyay, R., 2016.
\newblock ``Prevention of hot spot temperature in a distribution transformer
  using magnetic fluid as a coolant''.
\newblock {\em International Journal of Thermal Sciences, {\bf 103}}, May,
  pp.~35 -- 40.

\bibitem{Weinberg}
Weinberg, M., and Senyurt, A., 2017.
\newblock Polyamide electrical insulation for use in liquid filled
  transformers.

\bibitem{raus}
Lekawa‐Raus, A., Patmore, J., Kurzepa, L., Bulmer, J., and Koziol, K., 2014.
\newblock ``Electrical properties of carbon nanotube based fibers and their
  future use in electrical wiring''.
\newblock {\em Advanced Functional Materials, {\bf 24}}(24), March,
  pp.~3661--3682.

\bibitem{confPaper}
Esmailie, F., Cavilla, M.~S., and Ameel, T.~A., 2017.
\newblock ``A thermal transient model of heat transfer within an omnimagnet''.
\newblock In ASME Proceedings-Heat Transfer and Thermal Engineering,
  pp.~V008T10A046--V008T10A046--10.

\bibitem{Shampine}
Shampine, L., and Reichelt, M., 1997.
\newblock ``The matlab ode suite''.
\newblock {\em SIAM Journal on Scientific Computing, {\bf 18}}(1), November,
  pp.~1--22.

\bibitem{kiusalaas_2015}
Kiusalaas, J., 2015.
\newblock {\em Numerical Methods in Engineering with MATLAB®}, 3~ed.
\newblock Cambridge University Press.

\bibitem{Incropera}
Incropera, F.~P., DeWitt, D.~P., Lavine, A.~S., and Bergman, T.~L., 2011.
\newblock {\em Fundamentals of heat and mass transfer}, 7~ed.
\newblock Wiley.

\bibitem{cOPPER}
Pure copper.
\newblock http://www-ferp.ucsd.edu/LIB/PROPS/PANOS/cu.html.
\newblock Last access July 2020.

\bibitem{cOPPER2}
Haynes, W.~M., Lide, D.~R., and Thomas~J., B., 2016.
\newblock {\em Handbook of Chemistry and Physics: A Ready-reference Book of
  Chemical and Physical Data}.
\newblock Boca Raton, Florida : CRC Press.

\bibitem{copperelectricalresistivity}
Temperature coefficient of copper.
\newblock
  https://www.cirris.com/learning-center/general-testing/special-topics/177-temperature-coefficient-of-copper,
  Last access July 2020.
\newblock
  https://www.cirris.com/learning-center/general-testing/special-topics/177-temperature-coefficient-of-copper+.

\bibitem{copperelectricalresistivity2}
Dellinger, J.~H., 1970.
\newblock {\em Bulletin of the Bureau of Standards}.
\newblock Washington : U.S. G.P.O.

\bibitem{emissivity}
test, 2003.
\newblock Emissivity coefficients materials.
\newblock
  $https://www.engineeringtoolbox.com/emissivity-coefficients-d_447.htmll$,
  Last access July 2020.

\bibitem{Polyamide}
Polyamide - nylon 6 (pa 6).
\newblock http://www.goodfellow.com/E/Polyamide-Nylon-6.html.
\newblock Last access July 2020.

\bibitem{core}
{EFI} alloy 50 (aka magnifer 502, carpenter high permeability 491, alloy
  47-50).
\newblock
  https://edfagan.com/magnifer-50-high-permeability-49-alloy-4750-rod-coil-square-bar.php.
\newblock Last access July 2020.

\bibitem{coverins}
Polyurethane insulation-engineering toolbox.
\newblock
  {https://www.engineeringtoolbox.com/polyurethane-insulation$-k-values-d_1174$.html},
  Last access July 2020.

\bibitem{Catton}
Catton, I., 1978.
\newblock ``Natural convection in enclosures''.
\newblock {\em Proceeding 6th international heat transfer conference, {\bf 6}},
  pp.~13--31.

\bibitem{Churchill}
Churchill S.~W, H. H. S.~C., 1975.
\newblock ``Correlating equations for laminar and turbulent free convection
  from a vertical plate''.
\newblock {\em International journal of heat and mass transfer, {\bf 18}},
  pp.~1323--1329.

\bibitem{Radzeimska}
Radziemska, E., and Lewandowski, W., 2001.
\newblock ``Heat transfer by natural convection from an isothermal
  downward-facing round plate in unlimited space''.
\newblock {\em Applied Energy, {\bf 68}}(4), pp.~347 -- 366.

\bibitem{Liyod}
Lloyd, J.~R., and Moran, W.~R., 1974.
\newblock ``Natural convection adjacent to horizontal surface of various
  planforms''.
\newblock {\em Journal of Heat Transfer, {\bf 96}}(4), pp.~443--447.

\end{thebibliography}

\newpage
\setcounter{page}{1}
\section*{Appendix A: A Case Study}

The implementation of the lumped-capacitance transient model on the Omnimagnet used for the validation experiments is presented in this section.~The Omnimagnet consists of 16 components as indicated by the volume of each in Table \ref{Component volume}.  The components include the three solenoids, four frames, wire insulation on each solenoid (3), cover insulation on each solenoid (3), the air trapped within the inner solenoid by the middle solenoid (inner air), the spherical core, and the electric insulation paper between Solenoid 2 and Frame 4. Thin insulating paper is placed between all solenoids and their adjacent frames; however, their impact on heat transfer in negligible with one exception. The insulating paper between Frame 4 and Solenoid 2 is included, because it transmits radiation to the ambient. Applying Eq.~\ref{1} to all 16 elements produces a set of 16 ODE equations with 16 unknown temperatures, which are solved at each time step. The input data for the model are presented in Tables \ref{Component volume} to \ref{correction coefficients2}. Table \ref{Thermal properties} contains the constant thermal properties for the materials in each component. While the exact composition of the wire insulation is unknown, it consists primarily of polyamide. The specific heat capacity at constant pressure $c{_p}{_c}$  (Eq.~\ref{eq:2}\cite{Incropera,cOPPER,cOPPER2}) and thermal conductivity $\kappa_c$ (Eq.~\ref{eq:3}\cite{Incropera,cOPPER,cOPPER2}) of copper are modeled as functions of temperature.

Volume data for each component are presented in Table~\ref{Component volume}.~The volumes of the frames and core are extracted from SOLIDWORKS CAD files.~The volume of the solenoids, insulation layers, and the trapped air can be calculated using expressions provided in \cite{Petruska2014-2}.
\begin{table}[h]
\renewcommand\thetable{A.1} 
\caption{Component volume}
\begin{center}
\label{Component volume}
\begin{tabular}{| c | c|  l || c |c| l |} 
\hline
Variable & Element&  $V$~($cm\textsuperscript{3}$) &  Number & Element&  $V$~($cm\textsuperscript{3}$)\\
\hline
$V_1$& Solenoid 1 (Copper 1) &  326 & $V_9$ &Cover insulation 1 (Solenoid 1)&  4.2  \\
$V_2$   & Solenoid 2 (Copper 2) & 285 & $V_{10}$ &Wire insulation 2 (Solenoid 2)&  47.6 \\
$V_3$ & Solenoid 3 (Copper 3)  & 320 &$V_{11}$ & Cover insulation 2 (Solenoid 2) & 5.5 \\
$V_4$ &Frame 1  & 2$\times$27.77 & $V_{12}$ &Wire insulation 3 (Solenoid 3)&52.4 \\
$V_5$ &Frame 2 & 2$\times$31.95 &$V_{13}$ & Cover insulation 3 (Solenoid 3) & 6.9 \\
$V_6$&Frame 3  & 2$\times$25.46 & $V_{14}$ & Inside air&294\\
$V_7$&Frame 4  & 2$\times$46.18 & $V_{15}$&Spherical core & 525\\
$V_8$&Wire insulation 1 (Solenoid 1)  & 54.9 & $V_{16}$&Insulating paper & 2.9\\

\hline
\end{tabular}
\end{center}
\end{table}

\begin{equation}
 c{_p}{_c} = 316.21 + 0.3177~T{_c} -3.5\times10^{-4}~T{_c}^2 
 \label{eq:2}
 \tag{A.1}
\end{equation}

\begin{equation}
 \kappa{_c} = 420.75-0.068493~T{_c}
 \label{eq:3}
 \tag{A.2}
\end{equation}

The electrical resistance of wire is also modeled as a function of temperature (Eq.~\ref{eq:4}): 

\begin{equation}
 R = R{_0}(1+\alpha_0 (T-293.15) )
 \label{eq:4}
 \tag{A.3}
\end{equation}

where $R_{0}$ is the initial electrical resistance of the wire, which is 3.2~$\Omega$, 2.8~$\Omega$, and 3.1~$\Omega$ for Solenoids 1, 2, and 3, respectively. The constant $\alpha_0$ for copper is 0.0039 1/K \cite{copperelectricalresistivity,copperelectricalresistivity2}.

\begin{table}[h]
\renewcommand\thetable{A.2} 
\caption{Thermal properties for component materials}
\begin{center}
\label{Thermal properties}
\begin{tabular}{|c|| l| l| l| l|} 
\hline
Component & $c_p$~(kJ$\cdot$kg\textsuperscript{-1}$\cdot$K\textsuperscript{-1}) & $\rho$~(kg$\cdot$ m$\textsuperscript{-3}$) & $k$~(W$\cdot$m\textsuperscript{-1}$\cdot$K\textsuperscript{-1}) & $\epsilon$\\
\hline
Frame (aluminum) & 0.896~\cite{Incropera} & 2700~\cite{Incropera} & 210~\cite{Incropera} & 0.03~\cite{Incropera}\\
Wire (copper) & Eq.~\ref{2}~\cite{Incropera,cOPPER,cOPPER2} & 8960~\cite{Incropera} & Eq.~\ref{3}~\cite{Incropera,cOPPER,cOPPER2}& 0.78 (heated)~\cite{emissivity}\\
Wire insulation (polyamide) & 1.670~\cite{Polyamide} & 1140~\cite{Polyamide} & 0.25~\cite{Polyamide}&0.6 \\
Ferromagnetic core & 0.502~\cite{core} & 8165.57~\cite{core} & 13~\cite{core}&-\\
Cover insulation (polyurethane) & 1.800~\cite{coverins} & 1200~\cite{coverins} & 0.024~\cite{coverins}&0.9\\
Inside trapped air & 1.007~\cite{Incropera} & 1.614~\cite{Incropera} & 0.0263~\cite{Incropera} &- \\
Insulating paper & 1.336~\cite{Incropera} & 950~\cite{Incropera} & 0.143~\cite{Incropera} & 0.95~\cite{Incropera} \\
\hline
\end{tabular}
\end{center}
\end{table}

Calculating the contact surface area between two adjacent components (Table~\ref{Contact surface area}) is time consuming and prone to uncertainty. In addition, the contact surface area depends on the Cartesian direction, as conduction is a vector quantity. Thus, in addition to the contact surface area, component thickness is determined for all three directions, as shown in Table~\ref{component thickness}. The thickness of one layer of wire insulation, which is defined as $\Delta x_{17,m}$, is 0.004 cm. The thickness of wire insulation between frame 4 and wire insulation 3 is considered to be equal to the thickness of the wire insulation.


\begin{table}[]
\renewcommand\thetable{A.3} 
\setlength{\tabcolsep}{4pt}
\caption{Contact surface area}
\begin{center}
\label{Contact surface area}

\begin{tabular}{|l|l|c|c|c|} 
    \multicolumn{5}{c}{}\\
\hline
$S_{i,j,m}$& Element & $m$ = 1 &$m$ = 2&$m$ = 3 \\

 &  & $x$ direction (cm$^2$)&  $y$ direction (cm$^2$)& $z$ direction (cm$^2$) \\
\hline

\hline

$S_{1,8,m}$ & Solenoid 1 + Wire insulation 1 & 112 & 179 & 177 \\
$S_{1,9,m}$ & Solenoid 1 + Cover insulation 1 &  0 & 228& 232\\
$S_{2,10,m}$ & Solenoid 2 + Wire insulation 2 & 266 & 38 & 294 \\
$S_{2,11,m}$ &Solenoid 2 + Cover insulation 2& 326 & 0 & 285\\
$S_{3,12,m}$ & Solenoid 3 + Wire insulation 3& 381 & 362  & 72 \\
$S_{3,13,m}$ & Solenoid 3 + Cover insulation 3 & 422 & 403 & 0\\
$S_{4,5,m}$ & Frame 1 +  Frame 2 & 8 & 0 & 0 \\
$S_{4,8,m}$ & Frame 1 + Wire insulation 1 &0 & 95  &  20 \\ 
$S_{4,14,m}$ & Frame 1 + Trapped inside air& 11 & 90 & 12 \\
$S_{4,15,m}$ & Frame 1 + Sphere & 0 & 17 & 0 \\
$S_{5,6,m}$ & Frame 2 + Frame 3& 5 & 0 & 0 \\
$S_{5,8,m}$ & Frame 2 + Wire insulation 1 & 47 & 0 &  0  \\
$S_{5,10,m}$ & Frame 2 + Wire insulation 2& 138 & 0 &  14\\
$S_{5,12,m}$ & Frame 2 + Wire insulation 3 & 0 & 3.6 &  0  \\
$S_{5,14,m}$ & Frame 2 + Trapped inside air & 65 & 27 & 43 \\
$S_{6,0,m}$ & Frame 3 + Ambient air& 43 & 157 &  29.3 \\
$S_{6,7,m}$ & Frame 3 + Frame 4& 0 & 6 &  0 \\
$S_{6,10,m}$ & Frame 3 + Wire insulation 2 & 0 & 5.4 &  0 \\
$S_{6,12,m}$ & Frame 3 + Wire insulation 3 & 15 & 155 &  0 \\
$S_{7,0,m}$ & Frame 4 + Ambient air & 6 & 36 &  172  \\
$S_{7,12,m}$ & Frame 4 + Wire insulation 3 & 0 & 0 &  9 \\
$S_{7,16,m}$ & Frame 4 +Paper & 0 & 0 &  43\\
$S_{8,14,m}$ & Wire insulation 1 + Trapped inside air &  0 & 90 &  151\\
$S_{9,0,m}$ & Cover insulation 1 + Ambient air (middle channel)
& 0 & 134 &  0 \\
$S_{9,10,m}$ &  Cover insulation 1+Wire insulation 2  & 0 & 0 &  177 \\
$S_{10,14,m}$ & Wire insulation 2 + Inside air &  72 & 0 &  220\\
$S_{11,0,m}$ & Cover insulation 2 + Ambient air & 0 & 0 &  160\\
$S_{11, 12,m}$ &Cover insulation 2 + Wire insulation 3 & 242 & 0 &  0\\
$S_{11,16,m}$ & Cover insulation 2 + Paper & 0 & 0 &  180 \\
$S_{12,0,m}$ & Wire insulation 3 + Ambient air (middle channel)& 0&178 &0  \\
$S_{13,0,m}$ & Cover insulation 3 + Ambient air& 436& 577 &  36 \\
$S_{15,14,m}$ & Sphere + Inside air & 73& 73 &  73 \\
$S_{16,0,m}$ & Paper + Ambient air & 0 & 0 &  160\\

\hline
\end{tabular}
\end{center}
\end{table}

\begin{table}[h]
\setlength{\tabcolsep}{4pt}
\renewcommand\thetable{A.4} 
\caption{Component thickness}
\label{component thickness}
\begin{center}
\begin{tabular}{| c | c | c| c || c | c|c|c| } 
\hline
$\Delta$ x$_{j,m}$ & $m$ = 1 &$m$ = 2&$m$ = 3 &$\Delta$ x$_{j,m}$ &$m$ = 1&$m$ = 2&$m$ = 3 \\
& ~$x$~direction~(cm) & ~$y$~direction~(cm) &  ~$z$~direction~(cm) &  & ~$x$~direction~(cm) & ~$y$~direction~(cm) &  ~$z$~direction~(cm)\\

\hline
$\Delta$x$_{1,m}$  & 10 & 1.4 & 1.4 &
$\Delta$x$_{2,m}$   & 0.8 & 11.9 & 0.8 \\
$\Delta$x$_{3,m}$  & 0.7 & 12.1 & 0.7 &
$\Delta$x$_{4,m}$  & 10 & 0.6 & 2\\
$\Delta$x$_{5,m}$  & 0.5 & 12.7  & 1.1 &
$\Delta$x$_{6,m}$  &  1.1 & 0.3 & 1.2\\
$\Delta$x$_{7,m}$  & 1.2 & 14.3  &  0.6 &
$\Delta$x$_{8,m}$  & 0.6& 0.08 & 0.08  \\
$\Delta$x$_{9,m}$  & 10.06 & 0.02 & 0.02 &
$\Delta$x$_{10,m}$ & 0.05 &0.7 & 0.05  \\
$\Delta$x$_{11,m}$ & 0.02 & 12.11 & 0.02  &
$\Delta$x$_{12.m}$  & 0.04 & 0.04 & 0.8  \\
$\Delta$x$_{13,m}$ & 0.02 & 0.02 &  13.67 &
$\Delta$x$_{14,m}$ & 0 & 0 &  0 \\
$\Delta$x$_{15,m}$ & 4.5 & 3.7 &  4.5 &
$\Delta$x$_{16,m}$ & 10.06 & 7.51 &  0.01\\

\hline
\end{tabular}
\end{center}
\end{table}

Natural convection heat transfer coefficients are based on empirical correlations as presented in Table~\ref{Convection coefficient}. The locations where they are applied are: 1) between the inside trapped air and adjacent components, 2) between ambient air and Solenoid 3, 3) between ambient and the lower surface of Solenoid 2, and 4) between ambient and the upper surface of Solenoid 2.

\begin{table}[h]
\renewcommand\thetable{A.5} 
\caption{Convection coefficients}
\begin{center}
\label{Convection coefficient}
\begin{tabular}{|l| l |l |l| l|} 
\hline
$h_{i,j}$ & Nusselt number correlation & Characteristic length (m) &i & Reference\\
\hline
$(h_{i,14})_v$ &  $\overline{Nu}{_L} = 0.18 (\frac {Pr}{0.2+Pr}Ra{_L})^{0.29}$ & $L= y_{\mbox{\scriptsize{i}}}-y_{\mbox{\scriptsize{j}}}$& 4, 5, 8, 10, 15  & \cite{Catton} \\
$(h_{i,0})_{encl}$& $\overline{Nu}{_L} = [0.825+\frac {0.387Ra_{L}^{1/6}}{[1+(0.492/Pr)^{9/16}]^{8/27}}]^{2}$ & $L=L_i$ & 6, 7, 9, 12, 13&\cite{Churchill}\\
$(h_{i,0})_{hb}$&  $\overline{Nu}{_L} = 0.52Ra{_L}^{1/5}$ & $L= A_{s}/P$ & 11, 16&\cite{Radzeimska} \\
$(h_{i,0})_{ha}$& $\overline{Nu}{_L} = 0.54Ra{_L}^{1/4}$ & $L= A_{s}/P$ & 11, 16&\cite{Liyod}\\

\hline
\end{tabular}
\end{center}
\end{table}

$\overline{Nu}{_L}=\frac{h_{i,j}L}{\kappa_i}$, $Pr=\frac{\nu}{\alpha}$, and $Ra_L=\frac{g\beta(T_{s}-T{_\infty})L^3}{\nu\alpha}$ represent average Nusselt number, Prandtl number, and Rayleigh number, respectively. $\nu$ is kinematic viscosity (m\textsuperscript{2}$\cdot$s\textsuperscript{-1}),  $\alpha$ is thermal diffusivity (m\textsuperscript{2}$\cdot$s\textsuperscript{-1}), $g$ is gravity (m$\cdot$s\textsuperscript{-2}), $\beta$ is thermal expansion coefficient (K\textsuperscript{-1}), and $L$ is characteristic length (m). The subscript "s" represents the surface and subscript "$\infty$" represents ambient. $A{_s}$ is the surface area of the hot surface (m\textsuperscript{2}), and $P$ is the perimeter of the hot surface (m).

The glass temperature of the insulation covering the outside of each solenoid is 120$^{\circ}$C, which is the lowest melting point of all the materials in the entire system. Therefore, this temperature determined the upper limit for the experiments. All heating experiments were terminated when the insulation achieved a temperature of 115$^{\circ}$C or after a total time of 2\,hr, whichever occurred first. As a result, the maximum time that the Omnimagnet was powered depends on input current; higher input current leads to shorter safe operating time. Experimental data acquired from this process were used to validate simulation results. Initially, the differences in solenoid temperatures between the experiment and simulations was as high as 25$^{\circ}$C; however, the transient trends were similar. The maximum temperature difference was found to occur at the upper operational limits near 115$^{\circ}$C.
 
 To reduce the error, and to ultimately create a more accurate model, 15 coefficients were added to the basic code. These correction coefficients were applied to the terms with uncertain dimensions (e.g., thickness, contact area) and where there was a high probability that contact resistance may be present. In addition, three correction coefficients were added to the convection terms to account for the uncertainty in convection heat transfer coefficients. 
 An optimization process was performed to determine appropriate values for the coefficients based on minimization of the solenoid temperature differences between experiments and simulations. During the optimization process, transient solenoid temperatures for seven experiments, corresponding to the seven combinations of powered solenoids, were evaluated and compared to simulation data determined under similar conditions. The MATLAB optimization toolbox (using the fmincon interior point algorithm \cite{kiusalaas_2015}) was applied by varying the 15 coefficients using a quasi-Newton method.  
 The optimized correction coefficients were then applied to the terms shown in Tables ~\ref{correction coefficients} and ~\ref{correction coefficients2}. For this specific model $CF_{i,j,m}=CF_{j,i,m}$. For the conduction correction coefficients, smaller values suggest conduction between components is insignificant; whereas, values close to unity indicate more dominant conduction processes. Convection term correction coefficients suggest the heat transfer coefficients were underestimated by the correlations listed in Table~\ref{Convection coefficient}.

\begin{table}[h]
\renewcommand\thetable{A.6} 
\caption{Correction coefficients}
\begin{center}
\label{correction coefficients}
\begin{tabular}{|c |l |l |l |l|} 
\hline
 $CF_{i,j,m}$ & Component 1 & Component 2 &$m$ & value\\
\hline
$CF_{4,8,2}$ & Frame 1 & Wire insulation 1 & Conduction ($y$ direction)& 0.41\\
$CF_{4,8,3}$ & Frame 1 & Wire insulation 1 & Conduction ($z$ direction)&0.33\\
$CF_{4,15,2}$ & Frame 1 & Sphere & Conduction ($y$ direction)&0.04\\
$CF_{5,6,1}$ & Frame 2 & Frame 3 & Conduction ($x$ direction)&0.88\\
$CF_{5,8,1}$ & Frame 2 & Wire insulation 1 & Conduction ($x$ direction)&0.88\\
$CF_{5,10,1}$ & Frame 2 & Wire insulation 2 & Conduction ($x$ direction)& 0.13\\
$CF_{5,10,3}$ & Frame 2 & Wire insulation 2 & Conduction ($z$ direction)& 0.30\\
$CF_{5,12,3}$ & Frame 2 & Wire insulation 3& Conduction ($y$ direction)&0.60\\
$CF_{6,7,2}$ & Frame 3 & Frame 4 & Conduction ($y$ direction)&0.03\\
$CF_{6,10,2}$  & Frame 3 & Wire insulation 2 & Conduction ($y$ direction)&0.95\\
$CF_{6,12,1}$ & Frame 3 & Wire insulation 3 & Conduction ($x$ direction)&0.05\\
$CF_{6,12,2}$ & Frame 3 & Wire insulation 3 & Conduction ($y$ direction)&0.18\\
\hline
\end{tabular}
\end{center}
\end{table}

\begin{table}[h]
\renewcommand\thetable{A.7} 
\caption{Correction coefficients for convection coefficients}
\begin{center}
\label{correction coefficients2}
\begin{tabular}{|l |l |l |l |l|} 
\hline
 $CF_{i,j}$ & $i$ & Component 2 &value\\
\hline

($CF_{i,14})_{v}$ & 4, 5, 8, 10, 15 & Inside trapped air & 1.96 \\
($CF_{i,0})_{encl}$ & 6, 7, 9, 12, 13 & Ambient air  &  2.74\\
($CF_{i,0})_{ha,hb}$ & 11, 16 & Ambient air & 2.88 \\
\hline
\end{tabular}
\end{center}
\end{table} 


\section*{Appendix B: State-space Equations}

A state space equation may be used to predict the transient temperature through the entire Omnimagnet given a set of solenoid current values that may also vary with time.  The predictions may be compared with the corresponding temperatures recorded from a limited number of thermocouples embedded within the device. By applying the Kalman filter the error of prediction at each instance is reduced and temperatures at all points are calculated more precisely. An accurate prediction of temperature is desired to avoid increasing the temperature of any component above its melting point. Adjusting the current to avoid that scenario can be accomplished in real time using a feedback loop that includes temperature prediction. To this aim, the state space equation of an Omnimagnet is presented as Eq.~\ref{eq:5}:
\setlength{\belowdisplayskip}{0pt} \setlength{\belowdisplayshortskip}{0pt}
\setlength{\abovedisplayskip}{0pt} \setlength{\abovedisplayshortskip}{0pt}
\begin{equation}
 \dot{\vec{T}}=A{\vec{T}}+B{\vec{U}}+GT_{\mbox{\scriptsize\text{0}}}   
 \label{eq:5}
  \tag{B.1}
\end{equation}
where $\dot{\vec{T}}$ is a set of $n$ temperature derivatives with respect to time (for the Omnimagnet in Appendix A, $n$ = 16),  $A$ is an $n\times n$ matrix, $B$ is an $n\times3$ matrix, and $G$ is an $n\times1$ matrix. ${\vec{T}}$ is a vector consisting of $n$ temperatures, $\vec{U}$ is the input vector consisting of the square of three input currents ($I^2$). The non-zero components of the $A$, $B$, and $G$ matrices are listed here, in a general form and for the specific Omnimagnet presented in Appendix A. The round parenthesis are used when a parameter is function of temperature and brackets are used for grouping terms.

The general form of matrix $A$ components are defined by equations \ref{eq:6} and \ref{eq:7}:\\

\setlength{\belowdisplayskip}{0pt} \setlength{\belowdisplayshortskip}{0pt}
\setlength{\abovedisplayskip}{0pt} \setlength{\abovedisplayshortskip}{0pt}
\begin{equation}
A_{i,j} = \frac{1}{\rho_i c_{pi} V_i}\left[{\sum_{j=1,{j\neq i}}^{j= n}}\left(\sum_{m=1}^{m=3} \frac{\kappa_j S_{i,j,m}}{\Delta x_{j,m}}\right) +\left( \sum_{j=0|j=14}\sum_{m=1}^{m=3}h_{i,j}S_{i,j,m}\right) + \left(\sum_{m=1}^{m=3}\sigma\epsilon_i S_{i,0,m}T_i^3 \right)\right]
 \label{eq:6}
  \tag{B.2}
\end{equation}

\setlength{\belowdisplayskip}{0pt} \setlength{\belowdisplayshortskip}{0pt}
\setlength{\abovedisplayskip}{0pt} \setlength{\abovedisplayshortskip}{0pt}
\begin{equation}
A_{i,i} = -{\sum_{j=1,{j\neq i}}^{j= n}}A_{i,j}
 \label{eq:7}
  \tag{B.3}
\end{equation}\\

The matrix $A$ non-zero components for the Omnimagnet discussed on Appendix A, are:

\setlength{\belowdisplayskip}{0pt} \setlength{\belowdisplayshortskip}{0pt}
\setlength{\abovedisplayskip}{0pt} \setlength{\abovedisplayshortskip}{0pt}

\begin{equation*}
A_{1,1}=-[A_{1,8}+A_{1,9}],\quad
A_{1,8}=\frac{\kappa_8\left[\frac{S_{1,8,x}}{\Delta x_8}+\frac{S_{1,8,y}}{\Delta y_8}+\frac{S_{1,8,z}}{\Delta z_8}\right]}{\rho_1c_{p1}(T_1)V_1}, \quad 
A_{1,9}=\frac{\kappa_9\left[\frac{S_{1,9,x}}{\Delta x_9}+\frac{S_{1,9,y}}{\Delta y_9}+\frac{S_{1,9,z}}{\Delta z_9}\right]}{\rho_1c_{p1}(T_1)V_1}, \quad
\end{equation*}
\noindent
 --------------------------------------------------------------------------------------------------------------------------------------------------

\begin{equation*}
A_{2,2}=-[A_{2,10}+A_{2,11}], \quad
A_{2,10}=\frac{\kappa_{10}\left[\frac{S_{2,10,x}}{\Delta x_{10}}+\frac{S_{2,10,y}}{\Delta y_{10}}+\frac{S_{2,10,z}}{\Delta z_{10}}\right]}{\rho_2c_{p2}(T_2)V_2}, \nonumber \quad
A_{2,11}=\frac{-\kappa_{11}\left[\frac{S_{2,11,x}}{\Delta x_{11}}+\frac{S_{2,11,y}}{\Delta y_{11}}+\frac{S_{2,11,z}}{\Delta z_{11}}\right]}{\rho_2c_{p2}(T_2)V_2}, 
\end{equation*}
\noindent
--------------------------------------------------------------------------------------------------------------------------------------------------

\begin{equation*}
\noindent
A_{3,3}=-\left[A_{3,12}+A_{3,13}+\frac{\sigma\epsilon_{\scriptsize\text{3}}[S_{13,0,x}+S_{13,0,y}+S_{13,0,z}]T_{3}^3}{\rho_3c_{p3}(T_3)V_3}\right], \quad
A_{3,12}=\frac{\kappa_{12}\left[\frac{S_{3,12,x}}{\Delta x_{12}}+\frac{S_{3,12,y}}{\Delta y_{12}}+\frac{S_{3,12,z}}{\Delta z_{12}}\right]}{\rho_3c_{p3}(T_3)V_3},\quad
\end{equation*}

\begin{equation*}
A_{3,13}=\frac{\kappa_{13}\left[\frac{S_{3,13,x}}{\Delta x_{13}}+\frac{S_{3,13,y}}{\Delta y_{13}}+\frac{S_{3,13,z}}{\Delta z_{13}}\right]}{\rho_3c_{p3}(T_3)V_3},\nonumber\quad
\end{equation*}\\
\noindent
--------------------------------------------------------------------------------------------------------------------------------------------------

\begin{equation*}
\noindent
A_{4,4}=-\left[A_{4,5}+A_{4,8}+A_{4,14}+A_{4,15}\right],\quad
A_{4,5}=\frac{\kappa_{5}\left[\frac{S_{4,5,x}}{\Delta x_{5}}+\frac{S_{4,5,y}}{\Delta y_{5}}+\frac{S_{4,5,z}}{\Delta z_{5}}\right]}{\rho_4c_{p4}V_4}, \quad
A_{4,8}=\frac{\kappa_{8}\left[\frac{S_{4,8,x}}{\Delta x_{8}}+CF_{4,8,2}\frac{S_{4,8,y}}{\Delta y_{8}}+CF_{4,8,3}\frac{S_{4,8,z}}{\Delta z_{8}}\right]}{\rho_4c_{p4}V_4}, \nonumber\quad
\end{equation*}

\begin{equation*}
A_{4,14}=\frac{(CF_{4,14}h_{4,14})_v\left[S_{4,14,x}+S_{4,14,y}+S_{4,14,z}\right]}{\rho_4c_{p4}V_4}, \quad
A_{4,15}=\frac{\kappa_{15}\left[\frac{S_{4,15,x}}{\Delta x_{15}}+CF_{4,15,2}\frac{S_{4,15,y}}{\Delta y_{15}}+\frac{S_{4,15,z}}{\Delta z_{15}}\right]}{\rho_4c_{p4}V_4},
\end{equation*}\\
\noindent
--------------------------------------------------------------------------------------------------------------------------------------------------

\begin{equation*}
A_{5,4}=\frac{\kappa_{4}\left[\frac{S_{4,5,x}}{\Delta x_{4}}+\frac{S_{4,5,y}}{\Delta y_{4}}+\frac{S_{4,5,z}}{\Delta z_{4}}\right]}{\rho_5c_{p5}V_5},\nonumber\quad
A_{5,5}=-[A_{5,4}+A_{5,6}+A_{5,8}+A_{5,10}+A_{5,12}+A_{5,14}],\quad
A_{5,6}=\frac{\kappa_{6}\left[CF_{5,6,1}\frac{S_{5,6,x}}{\Delta x_{6}}+\frac{S_{5,6,y}}{\Delta y_{6}}+\frac{S_{5,6,z}}{\Delta z_{6}}\right]}{\rho_5c_{p5}V_5},\quad
\end{equation*}

\begin{equation*}
A_{5,8}=\frac{\kappa_{8}\left[CF_{5,8,1}\frac{S_{5,8,x}}{\Delta x_{8}}+\frac{S_{5,8,y}}{\Delta y_{8}}+\frac{S_{5,8,z}}{\Delta z_{8}}\right]}{\rho_5c_{p5}V_5}, \quad
A_{5,10}=\frac{\kappa_{10}\left[CF_{5,10,1}\frac{S_{5,10,x}}{\Delta x_{10}}+\frac{S_{5,10,y}}{\Delta y_{10}}+CF_{5,10,3}\frac{S_{5,10,z}}{\Delta z_{10}}\right]}{\rho_5c_{p5}V_5},\quad
\end{equation*}

\begin{equation*}
A_{5,12}=\frac{\kappa_{12}\left[\frac{S_{5,12,x}}{\Delta x_{12}}+\frac{S_{5,12,y}}{\Delta y_{12}}+CF_{5,12,3}\frac{S_{5,12,z}}{\Delta z_{12}}\right]}{\rho_5c_{p5}V_5},\quad
A_{5,14}=\frac{-(CF_{5,14}h_{5,14})_v[S_{5,14,1}+S_{5,14,2}+S_{5,14,3}]}{\rho_5c_{p5}V_5},
\end{equation*}\\
\noindent
--------------------------------------------------------------------------------------------------------------------------------------------------

\begin{equation*}
A_{6,5}=\frac{\kappa_{5}\left[CF_{5,6,1}\frac{S_{5,6,x}}{\Delta x_{5}}+CF_{5,6,2}\frac{S_{5,6,y}}{\Delta y_{5}}+\frac{S_{5,6,z}}{\Delta z_{5}}\right]}{\rho_6c_{p6}V_6},\nonumber\quad
A_{6,6}=-\left[A_{6,5}+A_{6,7}+A_{6,10}+A_{6,12}+\frac{(CF_{6,0}h_{6,0})_{encl}\left[S_{6,0,x}+S_{6,0,y}+S_{6,0,z}\right]}{\rho_6c_{p6}V_6}\right], \quad
\end{equation*}

\begin{equation*}
A_{6,7}=\frac{\kappa_{7}\left[\frac{S_{6,7,x}}{\Delta x_{7}}+\frac{S_{6,7,y}}{\Delta y_{7}}+\frac{S_{6,7,z}}{\Delta z_{7}}\right]}{\rho_6c_{p6}V_6},\quad
A_{6,10}=\frac{\kappa_{10}\left[\frac{S_{6,10,x}}{\Delta x_{10}}+CF_{6,10,2}\frac{S_{6,10,y}}{\Delta y_{10}}+\frac{S_{6,10,z}}{\Delta z_{10}}\right]}{\rho_6c_{p6}V_6},\quad
A_{6,12}=\frac{\kappa_{12}\left[CF_{6,12,1}\frac{S_{6,12,x}}{\Delta x_{12}}+CF_{6,12,2}\frac{S_{6,12,y}}{\Delta y_{12}}+\frac{S_{6,12,z}}{\Delta z_{12}}\right]}{\rho_6c_{p6}V_6},\nonumber
\end{equation*}\\
\noindent
--------------------------------------------------------------------------------------------------------------------------------------------------

\begin{equation*}
A_{7,6}=\frac{\kappa_{6}\left[\frac{S_{6,7,x}}{\Delta x_{6}}+CF_{6,7,2}\frac{S_{6,7,y}}{\Delta y_{6}}+\frac{S_{6,7,z}}{\Delta z_{6}}\right]}{\rho_7c_{p7}V_7}, \nonumber\quad
\end{equation*}

\begin{equation*}
A_{7,7}=-\left[A_{7,6}+A_{7,12}+A_{7,16}+\frac{-(CF_{7,0}h_{7,0})_{encl}\left[S_{7,0,x}+S_{7,0,y}+S_{7,0,z}\right]+\sigma\epsilon_{\scriptsize\text{7}}\left[S_{7,0,x}+S_{7,0,y}+S_{7,0,z}\right]T_{7}^3}{\rho_7c_{p7}V_7}\right], \quad
\end{equation*}

\begin{equation*}
A_{7,12}=\frac{\kappa_{12}\left[\frac{S_{7,12,x}}{\Delta x_{17}}+\frac{S_{7,12,y}}{\Delta y_{17}}+\frac{S_{7,12,z}}{\Delta z_{17}}\right]}{\rho_7c_{p7}V_7},\quad
A_{7,16}=\frac{\kappa_{16}\left[\frac{S_{7,16,x}}{\Delta x_{16}}+\frac{S_{7,16,y}}{\Delta y_{16}}+\frac{S_{7,16,z}}{\Delta z_{16}}\right]}{\rho_7c_{p7}V_7},\nonumber
\end{equation*}\\
\noindent
--------------------------------------------------------------------------------------------------------------------------------------------------

\begin{equation*}
A_{8,1}=\frac{\kappa_1(T_1)\left[\frac{S_{1,8,x}}{\Delta x_{1}}+\frac{S_{1,8,y}}{\Delta y_{1}}+\frac{S_{1,8,z}}{\Delta z_{1}}\right]}{\rho_8c_{p8}V_8},\nonumber\quad
A_{8,4}=\frac{\kappa_4\left[\frac{S_{4,8,x}}{\Delta x_{4}}+CF_{4,8,2}\frac{S_{4,8,y}}{\Delta y_{4}}+CF_{4,8,3}\frac{S_{4,8,z}}{\Delta z_{4}}\right]}{\rho_8 c_{p8} V_8},\quad
A_{8,5}=\frac{\kappa_5\left[CF_{5,8,1}\frac{S_{5,8,x}}{\Delta x_{5}}+\frac{S_{5,8,y}}{\Delta y_{5}}+\frac{S_{5,8,z}}{\Delta z_{5}}\right]}{\rho_8c_{p8}V_8},\quad
\end{equation*}

\begin{equation*}
A_{8,8}=-\left[A_{8,1}+A_{8,4}+A_{8,5}+A_{8,14}\right],\nonumber\quad
A_{8,14}=\frac{(CF_{8,14}h_{8,14})_v\left[S_{8,14,x}+S_{8,14,y}+S_{8,14,z}\right]}{\rho_8c_{p8}V_8},
\end{equation*}\\
\noindent
--------------------------------------------------------------------------------------------------------------------------------------------------

\begin{equation*}
A_{9,1}=\frac{\kappa_1(T_1)\left[\frac{S_{1,9,x}}{\Delta x_{1}}+\frac{S_{1,9,y}}{\Delta y_{1}}+\frac{S_{1,9,z}}{\Delta z_{1}}\right]}{\rho_9c_{p9}V_9},\quad
A_{9,9}=-\left[A_{9,1}+A_{9,10}+\frac{(CF_{9,0}h_{9,0})_{encl}\left[S_{9,0,x}+S_{9,0,y}+S_{9,0,z}\right]}{\rho_9c_{p9}V_9}\right],\quad
\end{equation*}

\begin{equation*}
A_{9,10}=\frac{\kappa_{10}\left[\frac{S_{9,10,x}}{\Delta x_{10}}+\frac{S_{9,10,y}}{\Delta y_{10}}+\frac{S_{9,10,z}}{\Delta z_{10}}\right]}{\rho_9c_{p9}V_9},\nonumber
\end{equation*}\\
\noindent
--------------------------------------------------------------------------------------------------------------------------------------------------

\begin{equation*}
A_{10,2}=\frac{\kappa_2(T_2)\left[\frac{S_{2,10,x}}{\Delta x_{2}}+\frac{S_{2,10,y}}{\Delta y_{2}}+\frac{S_{2,10,z}}{\Delta z_{2}}\right]}{\rho_{10}c_{p10}V_{10}},\quad
A_{10,5}=\frac{\kappa_5\left[CF_{5,10,1}\frac{S_{5,10,x}}{\Delta x_{5}}+\frac{S_{5,10,y}}{\Delta y_{5}}+CF_{5,10,3}\frac{S_{5,10,z}}{\Delta z_{5}}\right]}{\rho_{10}c_{p10}V_{10}},\nonumber\quad
\end{equation*}

\begin{equation*}
A_{10,6}=\frac{\kappa_6\left[\frac{S_{6,10,x}}{\Delta x_{6}}+CF_{6,10,2}\frac{S_{6,10,y}}{\Delta y_{6}}+\frac{S_{6,10,z}}{\Delta z_{6}}\right]}{\rho_{10}c_{p10}V_{10}},\quad
A_{10,9}=\frac{\kappa_9\left[\frac{S_{9,10,x}}{\Delta x_{9}}+\frac{S_{9,10,y}}{\Delta y_{9}}+\frac{S_{9,10,z}}{\Delta z_{9}}\right]}{\rho_{10}c_{p10}V_{10}},\quad
\end{equation*}

\begin{equation*}
A_{10,10}=-\left[A_{10,2}+A_{10,5}+A_{10,6}+A_{10,9}+A_{10,14}\right],\quad
A_{10,14}=\frac{(CF_{10,14}h_{10,14})_v\left[S_{10,14,x}+S_{10,14,y}+S_{10,14,z}\right]}{\rho_{10}c_{p10}V_{10}},\nonumber
\end{equation*}\\
\noindent
--------------------------------------------------------------------------------------------------------------------------------------------------

\begin{equation*}
A_{11,2}=\frac{\kappa_2(T_2)\left[\frac{S_{2,11,x}}{\Delta x_{2}}+\frac{S_{2,11,y}}{\Delta y_{2}}+\frac{S_{2,11,z}}{\Delta z_{2}}\right]}{\rho_{11}c_{p11}V_{11}},\quad
A_{11,11}=-\left[A_{11,2}+A_{11,12}+A_{11,16}+\frac{(CF_{11,0})_{ha,hb}((h_{11,0})_{hb}+(h_{11,0})_{ha})S_{11,0,y}}{\rho_{11}c_{p11}V_{11}}\right],\quad
\end{equation*}

\begin{equation*}
A_{11,12}=\frac{\kappa_{12}\left[\frac{S_{11,12,x}}{\Delta x_{12}}+\frac{S_{11,12,y}}{\Delta y_{12}}+\frac{S_{11,12,z}}{\Delta z_{12}}\right]}{\rho_{11}c_{p11}V_{11}},\nonumber\quad
A_{11,16}=\frac{\kappa_{16}\left[\frac{S_{11,16,x}}{\Delta x_{16}}+\frac{S_{11,16,y}}{\Delta y_{16}}+\frac{S_{11,16,z}}{\Delta z_{16}}\right]}{\rho_{11}c_{p11}V_{11}},
\end{equation*}\\
\noindent
\begin{equation*}
A_{12,3}=\frac{\kappa_3(T_3)\left[\frac{S_{3,12,x}}{\Delta x_{3}}+\frac{S_{3,12,y}}{\Delta y_{3}}+\frac{S_{3,12,z}}{\Delta
z_{3}}\right]}{\rho_{12}c_{p12}V_{12}},\nonumber\quad
A_{12,5}=\frac{\kappa_5\left[\frac{S_{5,12,x}}{\Delta x_{5}}+\frac{S_{5,12,y}}{\Delta y_{5}}+CF_{5,12,3}\frac{S_{5,12,z}}{\Delta
z_{5}}\right]}{\rho_{12}c_{p12}V_{12}},\quad
\end{equation*}

\begin{equation*}
A_{12,6}=\frac{\kappa_6\left[CF_{6,12,1}\frac{S_{6,12,x}}{\Delta x_{6}}+CF_{6,12,2}\frac{S_{6,12,y}}{\Delta y_{6}}+\frac{S_{6,12,z}}{\Delta
z_{6}}\right]}{\rho_{12}c_{p12}V_{12}},\quad
A_{12,7}=\frac{\kappa_7\left[\frac{S_{7,12,x}}{\Delta x_{7}}+\frac{S_{7,12,y}}{\Delta y_{7}}+\frac{S_{7,12,z}}{\Delta
z_{7}}\right]}{\rho_{12}c_{p12}V_{12}},\quad
A_{12,11}=\frac{\kappa_{11}\left[\frac{S_{11,12,x}}{\Delta x_{11}}+\frac{S_{11,12,y}}{\Delta y_{11}}+\frac{S_{11,12,z}}{\Delta
z_{11}}\right]}{\rho_{12}c_{p12}V_{12}}, \nonumber\quad
\end{equation*}

\begin{equation*}
A_{12,12}=-\left[A_{12,3}+A_{12,5}+A_{12,6}+A_{12,7}+A_{12,11}+\frac{\sigma\epsilon_{\scriptsize\text{12}}\left[S_{12,0,x}+S_{12,0,y}+S_{12,0,z}\right]T_{12}^3+(CF_{12,0}h_{12,0})_{encl}\left[S_{12,0,x}+S_{12,0,y}+S_{12,0,z}\right]}{\rho_{12}c_{p12}V_{12}}]\right],
\end{equation*}\\
\noindent
--------------------------------------------------------------------------------------------------------------------------------------------------

\begin{equation*}
A_{13,3}=\frac{\kappa_3(T_3)\left[\frac{S_{3,13,x}}{\Delta x_{3}}+\frac{S_{3,13,y}}{\Delta y_{3}}+\frac{S_{3,13,z}}{\Delta
z_{3}}\right]}{\rho_{13}c_{p13}V_{13}},\nonumber\quad
A_{13,13}=A_{13,3}+\frac{(CF_{13,0}h_{13,0})_{encl}\left[S_{13,0,x}+S_{13,0,y}+S_{13,0,z}\right]+\sigma\epsilon_{\scriptsize\text{13}}\left[S_{13,0,x}+S_{13,0,y}+S_{13,0,z}\right]T_{13}^3}{\rho_{13}c_{p13}V_{13}},
\end{equation*}\\
\noindent
--------------------------------------------------------------------------------------------------------------------------------------------------

\begin{equation*}
A_{14,4}=\frac{CF_{4,14}h_{4,14})_v\left[S_{4,14,x}+S_{4,14,y}+S_{4,14,z}\right]}{\rho_{14}c_{p14}V_{14}}, \quad
A_{14,5}=\frac{(CF_{5,14}h_{5,14})_v\left[S_{5,14,x}+S_{5,14,y}+S_{5,14,z}\right]}{\rho_{14}c_{p14}V_{14}},\nonumber
\end{equation*}

\begin{equation*}
\noindent
\it{A}_{14,8}=\frac{(CF_{8,14}h_{8,14})_v\left[S_{8,14,x}+S_{8,14,y}+S_{8,14,z}\right]}{\rho_{14}c_{p14}V_{14}},\\
A_{14,10}=\frac{(CF_{10,14}h_{10,14})_v\left[S_{10,14,x}+S_{10,14,y}+S_{10,14,z}\right]}{\rho_{14}c_{p14}V_{14}},\quad
\end{equation*}

\begin{equation*}
A_{14,14}=-\left[A_{14,4}+A_{14,5}+A_{14,8}+A_{14,10}+A_{14,15}\right],\quad
A_{14,15}=\frac{(CF_{15,14}h_{15,14})_v\left[S_{14,15,x}+S_{14,15,y}+S_{14,15,z}\right]}{\rho_{14}c_{p14}V_{14}},\nonumber
\end{equation*}\\
\noindent
--------------------------------------------------------------------------------------------------------------------------------------------------

\begin{equation*}
A_{15,4}=\frac{\kappa_4\left[\frac{S_{4,15,x}}{\Delta x_{4}}+CF_{4,15,2}\frac{S_{4,15,y}}{\Delta y_{4}}+\frac{S_{4,15,z}}{\Delta
z_{4}}\right]}{\rho_{15}c_{p15}V_{15}},\quad
A_{15,14}=\frac{(CF_{15,14}h_{15,14})_v\left[S_{14,15,x}+S_{14,15,y}+S_{14,15,z}\right]}{\rho_{15}c_{p15}V_{15}},\nonumber\quad
A_{15,15}=-\left[A_{15,4}+A_{15,14}\right],
\end{equation*}\\
\noindent
--------------------------------------------------------------------------------------------------------------------------------------------------

\begin{equation*}
A_{16,7}=\frac{\kappa_7\left[\frac{S_{7,16,x}}{\Delta x_{7}}+\frac{S_{7,16,y}}{\Delta y_{7}}+\frac{S_{7,16,z}}{\Delta
z_{7}}\right]}{\rho_{16}c_{p16}V_{16}},\quad
A_{16,11}=\frac{\kappa_{11}\left[\frac{S_{11,16,x}}{\Delta x_{11}}+\frac{S_{11,16,y}}{\Delta y_{11}}+\frac{S_{11,16,z}}{\Delta
z_{11}}\right]}{\rho_{16}c_{p16}V_{16}},
\quad
\end{equation*}

\begin{equation*}
A_{16,16}=-\left[A_{16,7}+A_{16,11}+\frac{(CF_{16,0})_{ha,hb}((h_{16,0})_{hb}+(h_{16,0})_{ha})\left[S_{16,0,x}+S_{16,0,y}+S_{16,0,z}\right]+\sigma\epsilon_{\scriptsize\text{16}}\left[S_{16,0,x}+S_{16,0,y}+S_{16,0,z}\right]T_{16}^3}{\rho_{16}c_{p16}V_{16}}\right]
\end{equation*}\\
\noindent
--------------------------------------------------------------------------------------------------------------------------------------------------

$B$ is an $n\times3$ matrix. The general form of the components of $B$ is (equation \ref{eq:8})
\setlength{\belowdisplayskip}{0pt} \setlength{\belowdisplayshortskip}{0pt}
\setlength{\abovedisplayskip}{0pt} \setlength{\abovedisplayshortskip}{0pt}
\begin{equation}
B_{i,j} = \frac{R_i}{\rho_i c_{pi} V_i}
 \label{eq:8}
  \tag{B.4}
\end{equation}\\
where the non-zero components represent the contribution of the electrical current.\\

For the Omnimagnet in Appendix A, $B$ is a $16\times3$ matrix and the non-zero components are:\\

\setlength{\belowdisplayskip}{0pt} \setlength{\belowdisplayshortskip}{0pt}
\setlength{\abovedisplayskip}{0pt} \setlength{\abovedisplayshortskip}{0pt}
\begin{eqnarray*}
B_{1,1}=\frac{R(T_1)}{\rho_1c_{p1}(T_1)V_1}\qquad
B_{2,2}=\frac{R(T_2)}{\rho_2c_{p2}(T_2)V_2}\qquad
B_{3,3}=\frac{R(T_3)}{\rho_3c_{p3}(T_3)V_3}\qquad
\end{eqnarray*}\\

$G$ is a $n\times1$ matrix in general and the components are expressed as (equation \ref{eq:9}):

\setlength{\belowdisplayskip}{0pt} \setlength{\belowdisplayshortskip}{0pt}
\setlength{\abovedisplayskip}{0pt} \setlength{\abovedisplayshortskip}{0pt}
\begin{equation}
G_{i,j} = \frac{1}{\rho_i c_{pi} V_i}\left[\left( {\sum_{m=1}^{m=3}h_{i,0}S_{i,0,m}}\right) + \left(\sum_{m=1}^{m=3}\sigma\epsilon_iS_{i,0,m}T_0^3 \right)\right]
 \label{eq:9}
  \tag{B.5}
\end{equation}\\

For the Omnimagnet in Appendix A, $G$ is a $16\times1$ matrix with the following nonzero components:
\setlength{\belowdisplayskip}{0pt} \setlength{\belowdisplayshortskip}{0pt}
\setlength{\abovedisplayskip}{0pt} \setlength{\abovedisplayshortskip}{0pt}
\begin{equation*}
G_3=\frac{\sigma\epsilon_{\scriptsize\text{3}}\left[S_{13,0,x}+S_{13,0,y}+S_{13,0,z}\right]T_{\scriptsize\text{0}}^3}{\rho_3c_{p3}(T_3)V_3},\quad
G_6=\frac{(CF_{6,0}h_{6,0})_{encl}\left[S_{6,0,x}+S_{6,0,y}+S_{6,0,z}\right]}{\rho_6c_{p6}V_6},\nonumber
\end{equation*}
\begin{equation*}
G_7=\frac{(CF_{7,0}h_{7,0})_{encl}\left[S_{7,0,x}+S_{7,0,y}+S_{7,0,z}\right]+\sigma\epsilon_{\scriptsize\text{7}}\left[S_{7,0,x}+S_{7,0,y}+S_{7,0,z}\right]T_{\scriptsize\text{0}}^3}{\rho_7c_{p7}V_7},\quad
\end{equation*}
\begin{equation*}
G_9=\frac{(CF_{9,0}h_{9,0})_{encl}\left[S_{9,0,x}+S_{9,0,y}+S_{9,0,z}\right]}{\rho_9c_{p9}V_9},\quad
G_{11}=\frac{(CF_{11,0})_{ha,hb}((h_{11,0})_{ha}+(h_{11,0})_{hb})S_{11,0,y}}{\rho_{11}c_{p11}V_{11}},\nonumber\quad
\end{equation*}
\begin{equation*}
G_{12}=\frac{(CF_{12,0}h_{12,0})_{encl}\left[S_{12,0,x}+S_{12,0,y}+S_{12,0,z}\right]+\sigma\epsilon_{\scriptsize\text{12}}\left[S_{12,0,x}+S_{12,0,y}+S_{12,0,z}\right]T_{\scriptsize\text{0}}^3}{\rho_{12}c_{p12}V_{12}},
\end{equation*}
\begin{equation*}
G_{13}=\frac{(CF_{13,0}h_{13,0})_{encl}\left[S_{13,0,x}+S_{13,0,y}+S_{13,0,z}\right]+\sigma\epsilon_{\scriptsize\text{13}}(\left[S_{13,0,x}+S_{13,0,y}+S_{13,0,z}\right]T_{\scriptsize\text{0}}^3}{\rho_{13}c_{p13}V_{13}},\nonumber\end{equation*}
\begin{equation*}
G_{16}=\frac{(CF_{16,0})_{ha,hb}((h_{16,0})_{ha}+(h_{16,0})_{hb})\left[S_{16,0,x}+S_{16,0,y}+S_{16,0,z}\right]+\sigma\epsilon_{\scriptsize\text{16}}\left[S_{16,0,x}+S_{16,0,y}+S_{16,0,z}\right]T_{\scriptsize\text{0}}^3}{\rho_{16}c_{p16}V_{16}}
\end{equation*}

\end{document}